\documentclass[aps,prb,epsfig,preprint,superscriptaddress,showpacs]{revtex4-2}
\usepackage{epstopdf}
\usepackage{graphicx}  
\usepackage{float}
\usepackage{dcolumn}   
\usepackage{bm}        
\usepackage{amssymb}
\usepackage[english]{babel}
\usepackage{lineno}

\begin{document}

\title{Pressure-induced lattice instabilities and phonon softening in the orthorhombically distorted ferrimagnet Ni$_4$Nb$_2$O$_9$}

\author{Rajesh Jana}
\email[Contact author: ]{rajesh.jana@hpstar.ac.cn}
\affiliation{Center for High Pressure Science and Technology Advanced Research (HPSTAR), Beijing 100193, P. R. China}
\affiliation{Department of Geological Sciences, Jackson School of Geosciences, The University of Texas at Austin, Austin, Texas 78712, USA}
\author{Xinyu Wang}
\affiliation{Center for High Pressure Science and Technology Advanced Research (HPSTAR), Beijing 100193, P. R. China}
\author{Takeshi Nakagawa}
\affiliation{Center for High Pressure Science and Technology Advanced Research (HPSTAR), Beijing 100193, P. R. China}
\author{Hirofumi Ishii}
\affiliation{Hirofumi Ishii, National Synchrotron Radiation Research Center, Hsinchu 30076, Taiwan}
\author{Alka Garg}
\affiliation{High Pressure and Synchrotron Radiation Physics Division, Bhabha Atomic Research Centre, Mumbai 400085, India}
\affiliation{Homi Bhabha National Institute, Anushaktinagar, Mumbai 400094, India}
\author{Rekha Rao}
\affiliation{Solid State Physics Division, Bhabha Atomic Research Centre, Mumbai 400085, India}
\affiliation{Homi Bhabha National Institute, Anushaktinagar, Mumbai 400094, India}
\author{Thomas Meier}
\email[Contact author: ]{thomasmeier@sharps.ac.cn}
\affiliation{Shanghai Key Laboratory MFree, Institute for Shanghai Advanced Research in Physical Sciences, Shanghai 201203, P. R. China.}
\affiliation{Center for High Pressure Science and Technology Advanced Research (HPSTAR), Beijing 100193, P. R. China}

\begin{abstract}

The ambient- and high-pressure behavior of the compensated ferrimagnet Ni$_4$Nb$_2$O$_9$, stabilized in an orthorhombically distorted honeycomb structure, is systematically investigated using nuclear magnetic resonance (NMR), Raman spectroscopy, and synchrotron x-ray diffraction. Ambient-pressure NMR measurements reveal that, despite its distinct orthorhombic symmetry, the local structural environment of Ni$_4$Nb$_2$O$_9$ closely resembles that of its trigonal analogue Mn$_4$Nb$_2$O$_9$. In contrast, the substantially different paramagnetic shifts observed in the two compounds reflect their distinct average crystal symmetries, which govern orbital overlap and magnetic exchange pathways. Under external pressure, Ni$_4$Nb$_2$O$_9$ exhibits pronounced sensitivity to lattice distortions and phonon instabilities. Three isostructural transitions are identified near 2.1, 6.2, and 9.9 GPa, manifested by Raman-mode splitting, anomalous frequency shifts, linewidth broadening, integrated-intensity anomalies, and slope changes in the pressure evolution of lattice parameters. At higher pressure, around 12.6 GPa, signatures of an incipient long-range structural transition from orthorhombic \textit{Pbcn} to monoclinic $P2/c$ symmetry emerge, signaling the onset of a symmetry-lowering transformation. The anomalous softening of the 191.5 cm$^{-1}$ Raman mode, accompanied by multiple linewidth and spectral-weight anomalies, serves as a key fingerprint of these structural instabilities, linking local symmetry breaking at low pressures to the long-range transition into the high-pressure $P2/c$ phase. Notably, pronounced linewidth anomalies, strongly anisotropic pressure coefficients spanning $+1.2$ to $-0.8$ cm$^{-1}$/GPa for low-frequency modes, together with a marked enhancement of the integrated intensity of the 137 cm$^{-1}$ low-frequency branch over the 2–12.6 GPa range, point toward a pressure-induced regime potentially influenced by coupled spin, orbital, and lattice degrees of freedom. The close correspondence of transition pressures in Ni$_4$Nb$_2$O$_9$ and those reported for Mn$_4$Nb$_2$O$_9$ highlights a common mechanism rooted in their similar local structural environments, as revealed by NMR.
\end{abstract}

\maketitle
\section{Introduction}

	The \textit{A}$_4$\textit{B}$_2$O$_9$ (\textit{A} = Fe, Co, Mn; \textit{B} = Nb, Ta) family has attracted significant interest due to its strong magnetoelectric and magneto-lattice couplings, making it an excellent platform for exploring the interplay among spin, lattice, and electronic degrees of freedom \cite{Ding, Panja, Khanh, Solovyev, Zheng, Narayanan, Khanh2}. The Mn-, Co-, and Fe-based members crystallize in a centrosymmetric trigonal \textit{P-3c1} structure (\textit{Z} = 2), hereafter referred to as the trigonal 429 systems, which are derived from the corundum framework  \cite{Bertaut, Jana}. In this structure, one cation site is replaced by Nb/Ta,  while the other splits into two inequivalent sites occupied by magnetic cations (\textit{A}1 and \textit{A}2). The unit cell comprises alternating nearly planar and buckled honeycomb layers stacked along the \textit{c} axis. In the planar layer, \textit{A}1O$_6$ octahedra shares edges with three neighboring \textit{A}1O$_6$ units, forming a two-dimensional honeycomb network in the \textit{ab} plane. The buckled layer contains two sublayers formed by alternating \textit{A}2O$_6$ and \textit{B}O$_6$ octahedra, which also form edge-sharing honeycomb motifs. These compounds exhibit pronounced magnetodielectric effects near their antiferromagnetic ordering temperature ($T_N$), spin-driven magnetoelectric coupling below $T_N$, and strong spin–phonon interactions arising from both short- and long-range magnetic correlations \cite{Park, Jana2, Jana3, Jana4}.

	In stark contrast, the Ni analogue Ni$_4$Nb$_2$O$_9$ (NNO) stabilizes in a lower-symmetry orthorhombic \textit{Pbcn} structure (\textit{Z} = 4), although it remains closely related to the trigonal arrangement \cite{Tailleur}. The structure is composed of two types of layers stacked along \textit{c} (Fig. 1). The planar honeycomb layer of edge-sharing Ni1O$_6$ octahedra resembles that of the trigonal 429 systems. However, the buckled layer is substantially different: Ni2O$_6$ octahedra connect to three NbO$_6$ and two Ni2O$_6$ units, forming zigzag ribbons running along the \textit{b} axis (Fig. 1(c)). Adjacent sublayers within the buckled layer are linked through corner-sharing Ni2O$_6$ octahedra, while the NbO$_6$ octahedra in two sublayers share their faces. Importantly, the orthorhombic structure hosts five inequivalent oxygen positions, resulting in five distinct Ni–O bond lengths, far more diverse than the two O sites and the comparatively uniform \textit{A}–O bond lengths found in the trigonal analogues. This increased diversity in Ni–O bond lengths endows the Ni–O network with enhanced structural flexibility, making NNO intrinsically more susceptible to external compression. As a result, NNO is expected to undergo more pronounced pressure-induced structural modifications than its trigonal counterparts, which can in turn promote stronger coupling between lattice distortions and other degrees of freedom, including magnetic and orbital order parameters.
	
	The magnetic behavior of NNO also departs fundamentally from the trigonal 429 systems. It undergoes a ferrimagnetic transition at ~76 K, and several studies report an absence of linear magnetoelectric or magnetodielectric effects at low temperatures. Nevertheless, NNO has drawn exceptional interest due to its unusual ferrimagnetism, including compensated magnetization and magnetic reversal below 35 K \cite{Fita, Zubov, Sannigrahi, Martin, Thota, Meng}. Typically, compensated ferrimagnetism arises from different magnetic ions or different valence states, but in NNO magnetic reversal occurs within a single magnetic species (Ni$^{2+}$) occupying two crystallographically inequivalent sites with distinct octahedral environments. This makes NNO a compelling candidate for magnetic switching, memory, and spintronic applications \cite{Fita, Meng}.
	
		In trigonal 429 systems, the two inequivalent magnetic moments (\textit{A}1 and \textit{A}2) are antiferromagnetically aligned within both honeycomb and buckled layers and are weakly ferromagnetically coupled along \textit{c}. In contrast, in orthorhombic NNO the Ni$^{2+}$ moments are ferromagnetically arranged within each layer, while successive layers couple antiferromagnetically. In this spin configuration, negative magnetization originates from the differing temperature dependences of the Ni1 and Ni2 sublattice moments \cite{Meng}. Recent work by Sannigrahi et al. demonstrated that distinct local distortions in the octahedral environments, along with differences in dimensionality and electronic structure of Ni1 and Ni2 sites, stabilize the ferrimagnetic ground state through unequal but opposite moment contributions \cite{Sannigrahi}. 	  
 
	Low-temperature Raman study on NNO reveals strong spin–phonon coupling below $T_C$, and pronounced magneto-lattice coupling around $T_C$ is manifested by a sharp contraction of the \textit{a} axis—perpendicular to the spin alignment, and anomalies in the dielectric constant \cite{Singh, Tailleur}. Although linear magnetoelectricity is symmetry-forbidden in orthorhombic NNO, the material exhibits technologically relevant magneto-lattice interactions, which can be tuned via external perturbations such as pressure. High-pressure Raman and x-ray diffraction (XRD) studies on trigonal 429 systems have uncovered rich sequences of pressure-induced structural and potentially magnetic transitions, including multiple isostructural distortions and several long-range transformations \cite{Jana2, Jana3, Jana5, Sahu}. In these systems, the emergence of magnetic interactions at high pressures is inferred from anomalous phonon linewidth behavior and activation of new Raman modes reminiscent of low-temperature responses. Notably, NMR studies on trigonal CNO and MNO show that the Mn-based analogue (MNO) possesses a substantially more distorted local environment than the Co-based compound, and Raman spectra exhibit similar trends \cite{Jana4}. Such differences in local structural distortion are decisive in setting the pressure scales for isostructural and long-range transitions, and they correlate strongly with the likelihood of pressure-induced magnetic interactions \cite{Jana5}.

 	Chemical substitution studies further highlight the structural control of magnetism in this family. In Ni$_{4-x}$Zn$_x$Nb$_2$O$_9$, $T_C$  decreases likely due to magnetic dilution, however remanent magnetization and compensation temperature increase up to $x=0.5$ before disappearing at $x=0.75$, indicating an important contribution from magnetocrystalline anisotropy and  lattice parameter modifications \cite{Bolletta}. Mg substitution in MgNi$_3$Nb$_2$O$_9$ also reduces $T_C$ from 76.5 to 45.5 K \cite{Tarakina}. Interestingly, magnetic cataion Co substitution in Ni$_{4-x}$Co$_x$Nb$_2$O$_9$ stabilizes the orthorhombic phase up to $x=2$, lowering $T_C$  from 77 K to 51 K, accompanied by unit-cell expansion and the disappearance of magnetic reversal above $x=0.17$ \cite{Jiongo-Dongmo}. Similar behavior occurs in NiCoZnNb$_2$O$_9$, further supporting a strong coupling between cation substitution, lattice metrics, and magnetic behavior \cite{Martin2}. Across these substitution series, the reductions in magnetic ordering temperatures correlate strongly with the increase in unit-cell volume and the decrease in the $b/a$ ratio, both of which modify the underlying superexchange geometry. Since external pressure drives the opposite effect with contracting the unit cell, it becomes essential to track the evolution of the $b/a$ ratio under compression and assess whether such structural tuning activates or modifies magnetic interactions at high pressure.

	In this work, we integrate Raman spectroscopy, nuclear magnetic resonance, and synchrotron x-ray diffraction to investigate the ambient and high-pressure behavior of orthorhombic NNO and to compare it with trigonal 429 analogues. NMR measurements reveal that the local environment in NNO closely resembles that of the Mn-based MNO, while differing substantially from Co-based CNO. Pressure induces strong and anisotropic modifications in both lattice and phonon behavior. Three isostructural transitions are identified from Raman mode splitting, anomalous phonon responses, and lattice-parameter evolution at low pressures. A long-range transition to the \textit{P2/c} phase emerges near 12.6 GPa in both Raman and XRD datasets. Additional lattice instabilities appear around 17.3 and 20.3 GPa, manifested in anomalies in lattice parameters, Raman frequencies, and linewidths. Notably, anomalous Raman linewidth and intensity behavior at the isostructural transitions, together with sharp increases in the $b/a$ ratio, suggest a potential reemergence of magnetic interactions at high pressures.     
	
	\begin{figure*}[ht!]
\includegraphics[width=14cm]{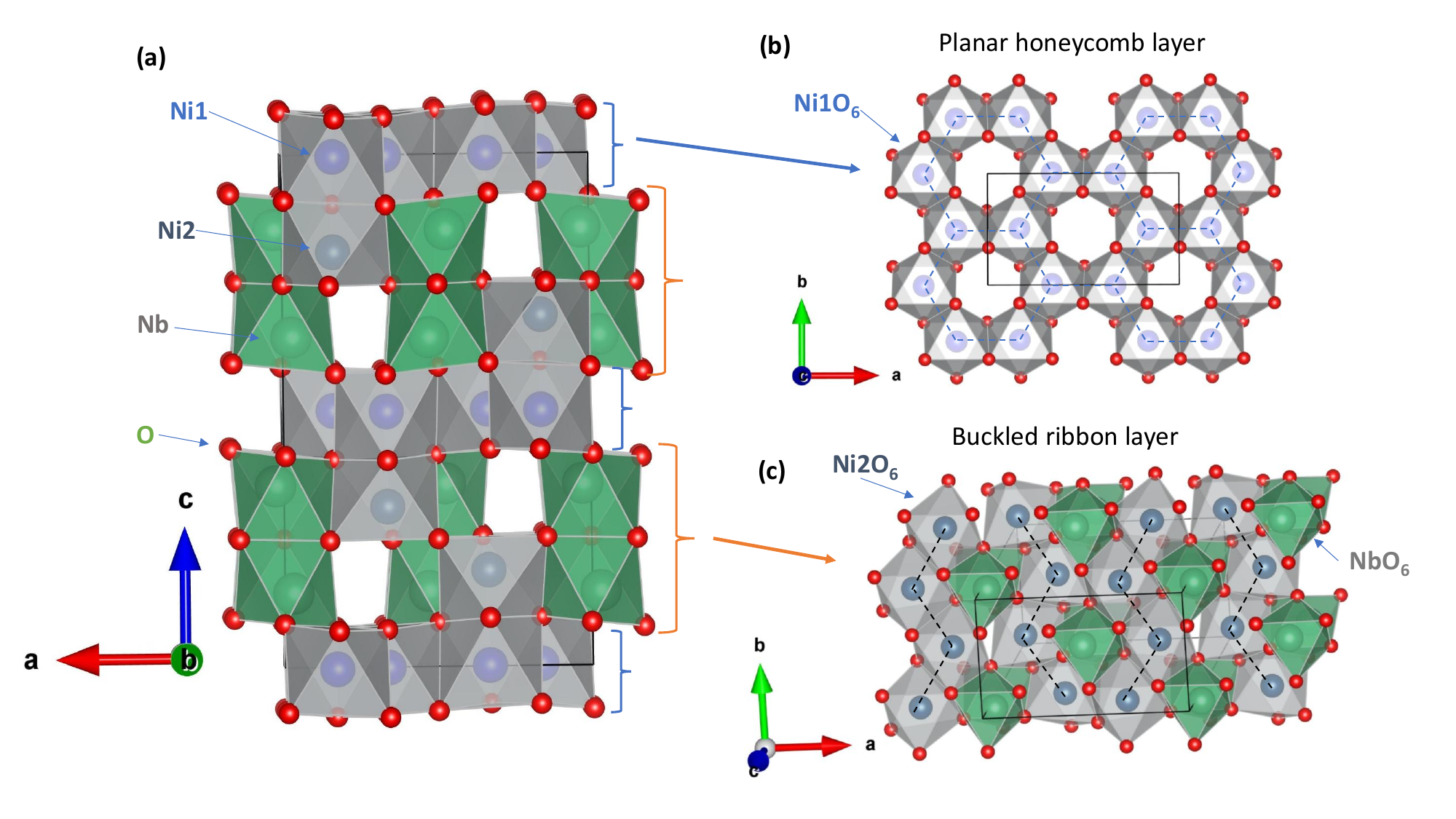}
\caption{\label{fig1}(Colour online) (a) Illustration of the unit cell of NNO in orthorhombic \textit{Pbcn} symmetry at ambient conditions.(b) Honeycomb arrangments of Ni1O$_6$ octahedra in the nearly planar layer viwed through c-axis. (c) Buckled layer made of Ni2O$_6$ ribbons running along the b axis and NbO$_6$ octahedra.}  
\end{figure*}

\section{Experimental and computational details}

\subsection{Sample synthesis and phase characterization}
High-quality polycrystalline NNO sample is prepared using a solid-state calcination method. Stoichiometric amounts of NiO and Nb$_2$O$_5$ are preheated at 300$^{\circ}$C for 3 h to remove moisture from the starting reagents, then thoroughly mixed using an agate mortar and pestle for several hours. The well-mixed powder was pelletized into discs and calcined at 1200$^{\circ}$C for 6 h in air. After the first calcination, the pellets are reground and repelletized for a second calcination at 1300$^{\circ}$C for 6 h. The final product is calcined at 1400$^{\circ}$C several times with intermediate grinding and pelletization to improve sample homogeneity. The ambient phase characterization is checked by rotating anode based x-ray diffractometer with Cu K$_{\alpha}$$_{1,2}$ radiation.

\subsection{Nuclear magnetic resonance specroscopy}

 Nuclear magnetic resonance (NMR) experiments on the $^{93}$Nb nucleus in NNO are performed at ambient conditions using a 9.3 T static superconducting magnet. A customized NMR probe with controllable tuning and matching capacitance, equipped with a home-made 5-turn Cu coil, is employed to optimize the radio-frequency excitation and detect the $^{93}$Nb signal within a narrow frequency window of 0.4 MHz. The complete spectrum is obtained by collecting signals at several central frequency spans covering the 92–102 MHz range. The spin–lattice relaxation time (T1) and spin–spin relaxation time (T2) are determined using solid-echo magnetic saturation recovery and magnetization decay measurements, respectively.

\subsection{High pressure Raman spectroscopy}
Ambient and high-pressure Raman scattering experiments up to 38.3 are carried out using a Renishaw inVia Raman microscope equipped with a 2400 g/mm grating and a 532 nm excitation laser. The scattered Raman signal is collected in a back-scattering geometry using a  long-working-distance, infinity-corrected 20x objective. High pressure is generated using a symmetric diamond anvil cell (DAC) with 300 $\mu$m ~culets. The sample, along with a ruby sphere for pressure calibration, is loaded into a sample cavity of 150 $\mu$m diameter and 30 $\mu$m thickness, prepared by pre-indenting a 200 $\mu$m-thick Re gasket and drilling a central hole using an electric discharge machine. Neon gas is used as the pressure-transmitting medium to maintain hydrostatic conditions inside the sample chamber. The pressure inside the sample chamber is accurately determined using the ruby fluorescence method \cite{Mao}.

\subsection{High-pressure synchrotron x-ray diffraction}
High-pressure synchrotron powder x-ray diffraction measurements are carried out in the pressure range of 4.3–34 GPa using a BX90-type DAC equipped with 300 $\mu$m culets at the SPring-8 facility (BL12B2, Hyogo, Japan). A monochromatic x-ray beam of wavelength 0.6199 ~\AA and beam size of 50 $\mu$m is employed for the experiments. The sample chamber is prepared by pre-indenting a Re gasket to a thickness of ~30 $\mu$m and drilling a central hole of 120 $\mu$m in diameter. Fine polycrystalline powder, together with small ruby chips used for pressure calibration, are loaded into the chamber. Silicone oil is used as the pressure-transmitting medium to ensure quasi-hydrostatic conditions throughout the pressure range. Calibration of the sample–detector distance, detector geometry, and conversion of the two-dimensional diffraction images into one-dimensional intensity versus 2$\theta$ profiles are performed using the Dioptas software package \cite{Prescher}. The resulting diffraction patterns are indexed using Crysfire/DICVOL and Checkcell \cite{Shirley, Boultif, Laugier}; and structural refinements are carried out using GSAS-EXPGUI/FullProf suite \cite{Toby, Carvajal}.

\section{Results and Discussion}

\subsection{Ambient phase characterization}

	The ambient XRD pattern of NNO, collected using Cu K$\alpha_{1,2}$ radiation, is shown in Fig. S1 of the Supplemental Material (SM) \cite{SM}. The formation of an almost single-phase sample is confirmed by the excellent fit of all diffraction peaks to the orthorhombic \textit{Pbcn} symmetry, except for a very weak peak (marked with *), which has also been reported previously \cite{Meng}. The Rietveld refinement yields lattice parameters $a = 8.7188(4)$, $b = 5.0725(6)$, and $c = 14.2921(5)$ ~\AA, in close agreement with earlier reports \cite{Tailleur}. The corresponding atomic coordinates are provided in Table S I \cite{SM}.

Group-theoretical analysis for the \textit{Pbcn} structure of NNO predicts 90 Raman-active modes ($22A_g + 23B_{1g} + 22B_{2g} + 23B_{3g}$) for four formula units per unit cell (60 atoms) \cite{Singh}. Experimentally, however, only 40 Raman modes are resolved at ambient conditions. The deconvolution of the Raman spectrum is shown in Fig. S2, where the observed modes are marked by blue arrows and labeled M(1)–M(41) for descriptive purposes \cite{SM}. A similar Raman spectrum at ambient temperature was reported in earlier low-temperature Raman study \cite{Singh}. The substantially larger number of observed Raman modes compared to the trigonal 429 systems \cite{Jana4} reflects the lower average crystal symmetry of orthorhombic NNO and the larger number of crystallographically inequivalent atomic sites, which relax selection rules and activate additional Raman modes.

\subsection{Nuclear magnetic resonance study}

Figure 2(a) presents the ambient-pressure $^{93}$Nb NMR spectrum of NNO recorded using a frequency-sweep protocol over the range 91–103 MHz. The spectrum is plotted on a relative frequency scale ($f - f_0$), where $f_0$ is the Larmor frequency of the nonmagnetic reference compound potassium hexachloroniobate(IV). The spectrum spans a broad frequency window of approximately 11 MHz and consists of a central transition centered at $f - f_0 \approx 338$ kHz (paramagnetic shift), together with four well-resolved quadrupolar satellite peaks, as expected for the $^{93}$Nb nucleus with nuclear spin $I = 9/2$. From the separation between adjacent quadrupolar satellite peaks, the quadrupolar frequency ($\delta_q$) is estimated to be approximately 860 kHz, reflecting a sizable electric field gradient at the Nb site.

\begin{figure*}[ht!]
\includegraphics[width=18cm]{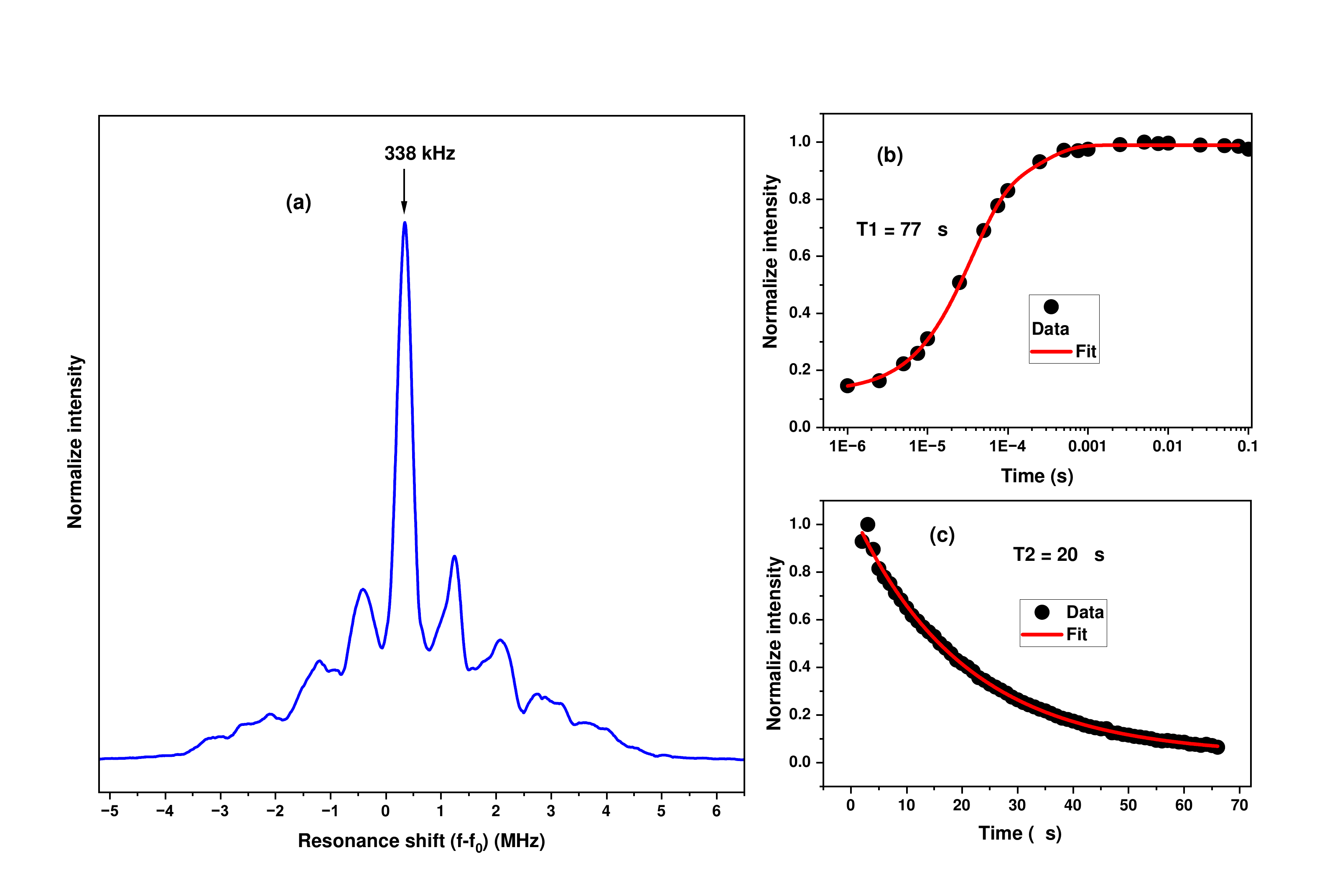}
\caption{\label{fig1}(Colour online) (a) $^{93}$Nb NMR spectrum of NNO at ambient conditions. (b) Spin-lattice relaxation constant (T1), measured using saturation recovery method. (c) Spin-spin relaxation constant (T2). Black cicles represent the experimental data, while red curves are the fit using the Eqs. 1 and 2 for saturation recovery and magnetization decay model, respectively. } 
\end{figure*}

Figure 2(b) shows the recovery of the longitudinal magnetization measured at variable delay times ranging from 1 $\mu$s to 100 ms. The spin–lattice relaxation constant $T1$ is extracted by fitting the data using the magnetic saturation-recovery expression

\begin{equation}
M_z = 1 - B\left[Ce^{-t/t_{11}} + (1-C)e^{-t/t_{12}}\right],
\end{equation}

where $B$ and $C$ are fitting constants, and $t_{11}$ and $t_{12}$ are characteristic time parameters that define $T1$. The best fit yields a spin–lattice relaxation time of $T1 = 77~\mu$s.

The spin–spin relaxation constant $T2$ is obtained from the decay of the transverse magnetization measured at variable delay times, as shown in Fig. 2(c). The data are fitted using a single-exponential decay function,

\begin{equation}
M_{xy} = y_0 + e^{-t/t_2},
\end{equation}

where $y_0$ is a constant offset and $t_2$ is the decay constant corresponding to $T2$, defined as the time at which the magnetization decays to $1/e$ (37\%) of its initial value. The fit yields $T2 = 20~\mu$s.

Table~I summarizes the paramagnetic shift, $T1$, $T2$, and quadrupolar frequency for NNO and compares them with those of the Mn- and Co-based analogues \cite{Jana4}. The significantly reduced paramagnetic shift in NNO compared to its trigonal analogues \cite{Jana4} (Table I) indicates a weaker hyperfine field at the Nb site, consistent with differences in the local magnetic environment arising from the lower magnetic susceptibility of Ni$^{2+}$ in NNO and the modified orbital overlap imposed by the orthorhombic crystal symmetry.  Notably, the $T1$ and $T2$ values of NNO are close to those of MNO but differ markedly from those of CNO, for which $T1$ is substantially longer. This behavior indicates that, despite its orthorhombic symmetry, the local structural environment of NNO more closely resembles that of MNO than CNO. This conclusion is further supported by the nearly identical quadrupolar frequencies observed for NNO and MNO, which are larger than that of CNO, pointing to similar local electric-field gradients at the Nb site.

				\begin{table*}[hbt]
\caption{\label{tab:table2
}
Comparison of NMR properties of NNO with MNO and CNO \cite{Jana4}.}
\begin{ruledtabular}
\begin{tabular}{ccccccc}
 Material  & Paramagnetic shift (kHz) & T1 ($\mu$s) & T2 ($\mu$s) & $\delta_q$ (kHz) \\
\hline
NNO        & 338 &   77 & 20 & 860 \\
\hline
 MNO       & 739 &  61  & 13 & 870\\
\hline
CNO        & 846 &  588 & 40 & 720 \\

\end{tabular}
\end{ruledtabular}
\end{table*}

\subsection{High pressure Raman studies on NNO}

Pressure-dependent Raman spectra up to 12.6 GPa, at selected representative pressures, are shown in Fig. 2. Under compression, nearly all Raman peaks harden, with the notable exception of the 400 cm$^{-1}$ mode, which shows a clear redshift. However, around 2.1 GPa, several notable changes are observed. A low-frequency mode at 138 cm$^{-1}$ (M(7)) splits into two distinct features (indicated by the upward red arrow in Fig. 2), M(7)$_{low}$ and M(7)$_{high}$, whose markedly different pressure responses cause them to diverge at higher pressures. Three modes at 206, 280 and 400 cm$^{-1}$ are completely suppressed at this pressure, as marked by the downward red arrows. With further compression to 6.2 GPa, the 129 cm$^{-1}$ M(6) mode gradually shifts to 130.7 cm$^{-1}$ and then splits into two modes at 129.1 cm$^{-1}$ (M(6)$_{low}$) and 132.7 cm$^{-1}$ (M(6)$_{high}$). At higher pressures, the M(6)$_{high}$ mode exhibits a sluggish blue shift, whereas in marked contrast, the M(6)$_{low}$ mode undergoes gradual softening, as highlighted by the dashed trend lines in Fig. 2. At the same pressure (6.2 GPa), two additional Raman modes appear at 242 ($\Omega$(5)) and 365 cm$^{-1}$ ($\Omega$(7)), accompanied by the disappearance of the M(30) and M(32) modes, indicated by the upward and downward arrows, respectively. Similar emergence/splitting and disappearance of Raman modes are also observed at 9.9 GPa.
	 
\begin{figure*}[ht!]
\includegraphics[width=18cm]{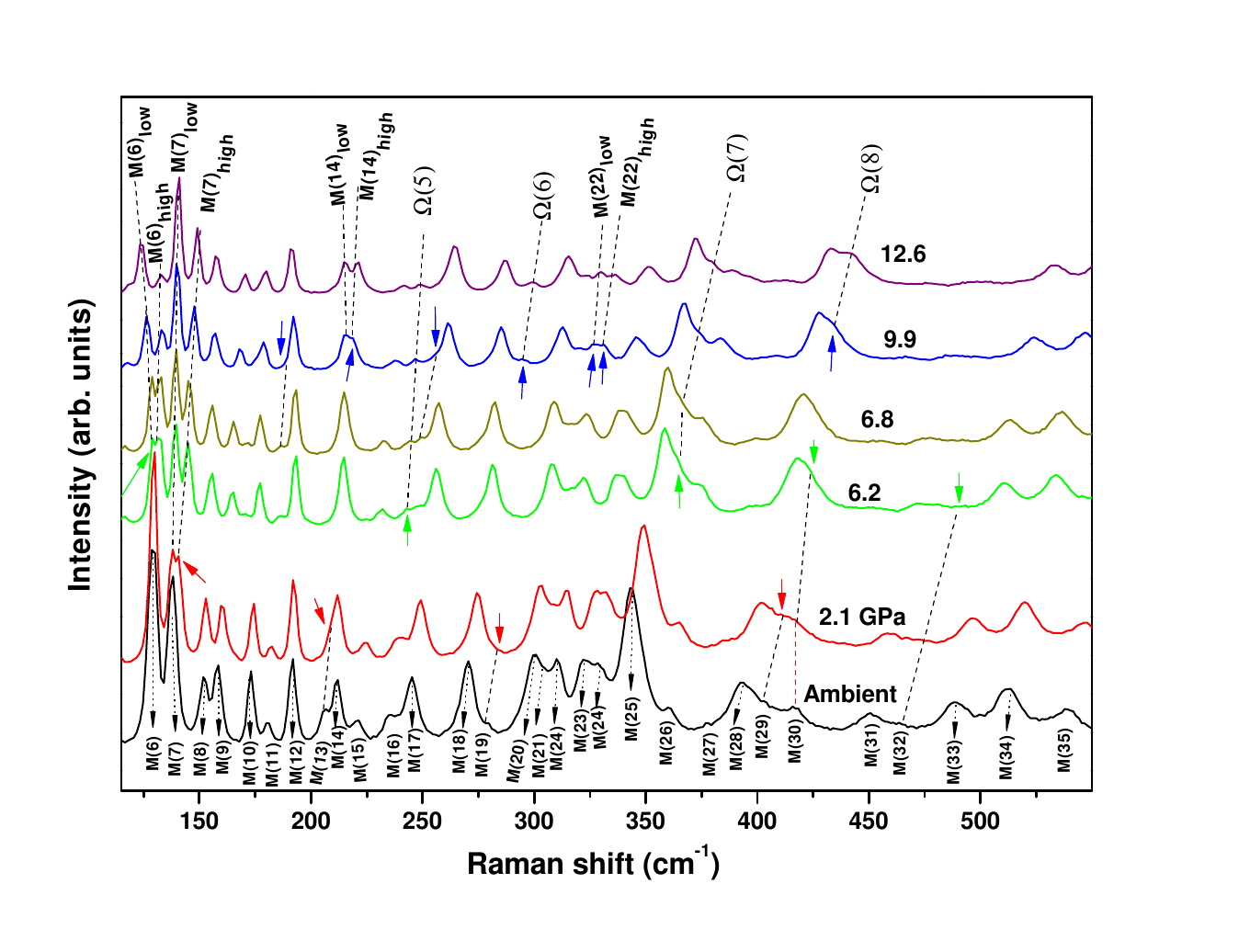}
\caption{\label{fig2}(Colour online) High pressure Raamn spectra of NNO up to 12.6 GPa. Splitting/appearance of Raman modes is highlighted by upward solid arrows, while disappearance of modes is indicated by downward solid arrows. Dashed lines follow the pressure evolution of these modes.} 
\end{figure*}

	At 12.6 GPa, a new octahedral Raman mode appears at around 789 cm$^{-1}$, denoted as $\omega$(26) in Fig. 4, signaling the onset of a long-range structural transition into a high-pressure (HP) phase, consistent with observations in trigonal 429 counterparts \cite{Jana2, Jana3, Jana5}. Upon further compression to 13.6 GPa, three additional Raman modes emerge (marked by the blue arrows in Fig. 4), indicating the progressive development of the HP phase. A pronounced reorganization of the Raman spectra is observed over the pressure range 17.3–20.5 GPa, marked by the appearance and disappearance of several modes, as highlighted by the upward and downward arrows in Fig. 4. In this pressure window, the intensity of the low-frequency 83 cm$^{-1}$ mode is strongly enhanced, and above 20.5 GPa the octahedral mode $\omega$(26) exhibits a rapid increase in intensity. Together, these spectral features reflect the rapid growth of the HP phase, accompanied by a strengthening of octahedral distortions, and point to a pressure-driven, pronounced structural rearrangement in Ni$_4$Nb$_2$O$_9$.

\begin{figure*}[ht!]
\includegraphics[width=18cm]{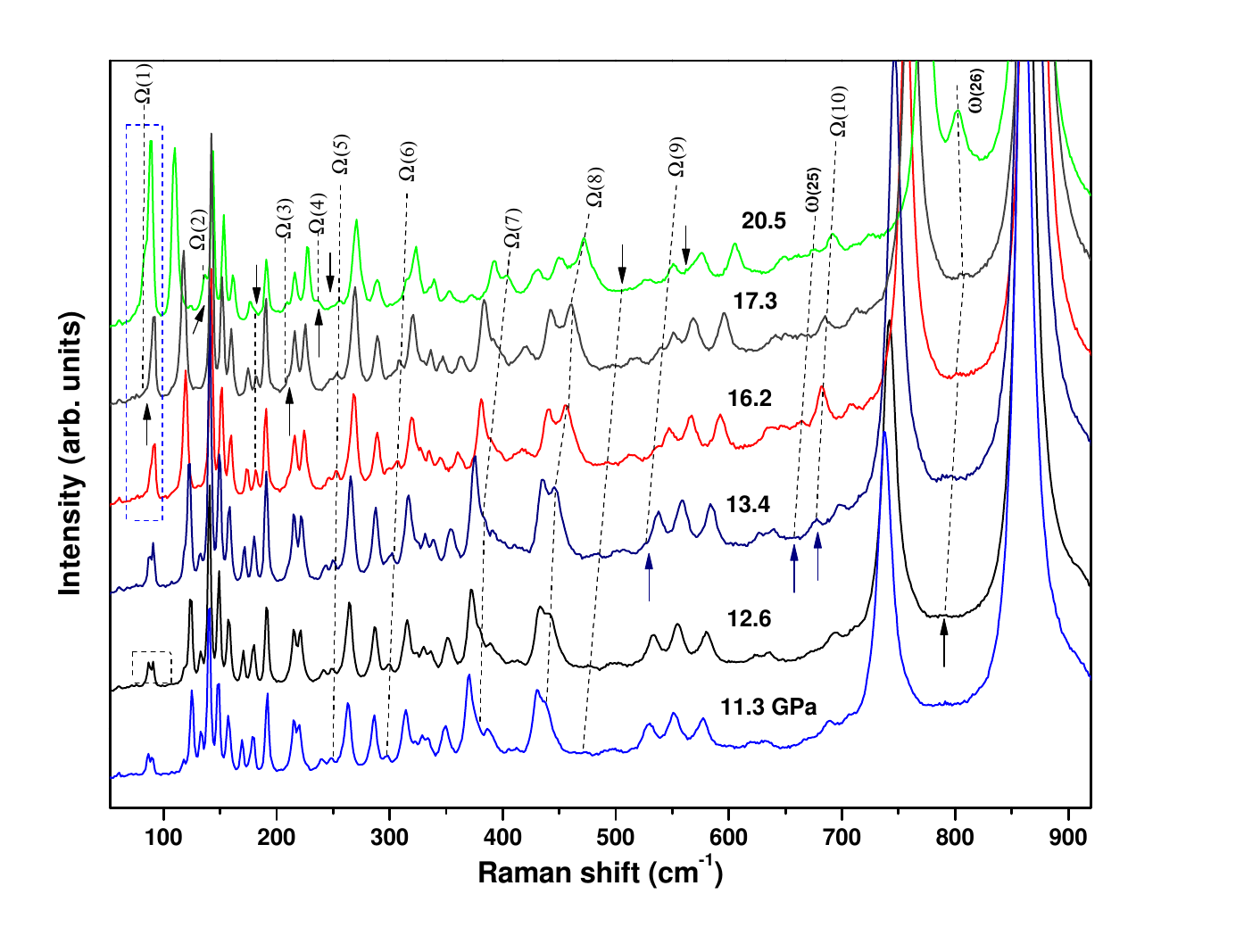}
\caption{\label{fig3}(Colour online) Higher pressure Raman spectra of NNO over the pressure range of 11.1-20.5 GPa. Upward and downard arrows marked the appearance and disappearance of modes, respectively, while their chnages with pressure is highlighted by the black dahsed lines.} 
\end{figure*}

	Raman spectra collected at higher pressures, up to 38.3 GPa, the maximum pressure explored in this study are shown in Fig. 5. In the pressure range 20.5–23 GPa, the spectra undergo extensive modifications. At 21.8 GPa, three new Raman modes emerge at approximately 202 ($\omega$(8)), 278 ($\omega$(12)), and 365 cm$^{-1}$ ($\omega$(24)), while five modes are completely suppressed, as indicated by the upward and downward arrows in Fig. 5(a), respectively. Upon further compression to 23 GPa, five additional modes below 380 cm$^{-1}$ and one mode near 611 cm$^{-1}$ disappear. Concurrently, the Raman modes in the 380–530 cm$^{-1}$ range experience substantial reorganization, evolving into distinctly different modes at 23 GPa compared with those observed at 20.5 GPa. Within this pressure interval, the octahedral mode $\omega$(26) associated with the high-pressure phase shows a pronounced enhancement in intensity, whereas the strongest Raman mode of the ambient \textit{Pbcn} phase rapidly diminishes. Moreover, at 23 GPa an additional octahedral Raman mode appears near 840 cm$^{-1}$, as labeled by $\omega$(27) in Fig. 5(b), providing further evidence of the evolving high-pressure structural landscape.

 \begin{figure*}[ht!]
\includegraphics[width=18cm]{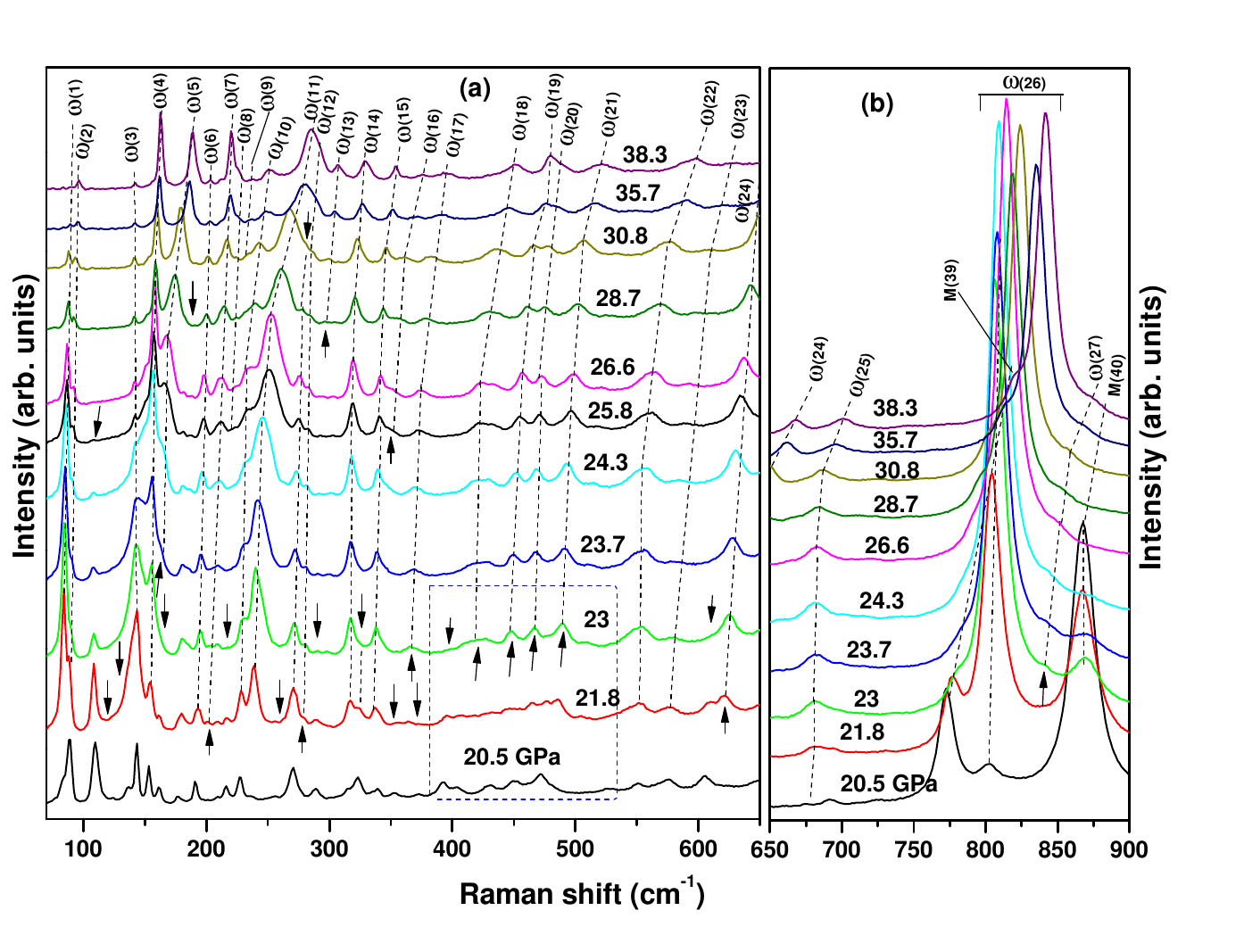}
\caption{\label{fig4} Raman spectra of NNO in the pressure range of 20.5-38.3 GPa for the wavenuber range (a) 60-650 cm-1 and (b) 650-900 cm-1. The appearance/disappearance of Raman modes indicate by the upward/downward arrows. Pressure evolution of each Raman modes is followed by the black dahsed lines. The blue dashed rectangle highlight the wavenumber range where Raman modes heavily renormalized and trnsformed into new modes in the pressre ramge 20.5-23 GPa.} 
\end{figure*}

	Beyond 23 GPa and up to 38.3 GPa, comparatively fewer changes are observed in terms of the appearance or disappearance of Raman modes. A new mode emerges at 351 cm$^{-1}$ ($\omega$(16)) at 25.8 GPa and another at 297 cm$^{-1}$ ($\omega$(13)) at 28.7 GPa, while at each of these pressures one mode is suppressed, as indicated by the downward arrows. Notably, the strongest Raman mode M(40 )associated with the ambient \textit{Pbcn} phase is completely suppressed above 26 GPa. Nevertheless, a continuous evolution of the peak shapes of several Raman modes persists up to the highest pressure investigated. This progressive renormalization results in the stabilization of three sharp modes at approximately 163, 189, and 220 cm$^{-1}$, labeled by $\omega$(4), $\omega$(5), and $\omega$(7), respectively at 38.3 GPa.

	In Fig. 6, the pressure evolution of the frequency, linewidth and integrated intensity of the two low-frequency modes M(3), M(4), and M(7) is shown up to 22 GPa, beyond which the high-pressure (HP) phase becomes dominant. The M(4) mode remains nearly pressure-independent up to about 2 GPa and merges with the M(3) mode at 2.1 GPa (Fig. 6(a)). The M(3) mode, in contrast, gradually hardens and exhibits a distinct slope change near 2.1 GPa. Above 2.8 GPa, the two modes become clearly distinguishable again, and the M(3) mode continues to harden up to 17.3 GPa, showing additional slope changes at approximately 6.2, 9.9, and 12.6 GPa. Beyond 17.3 GPa, it undergoes rapid softening up to 20.3 GPa, followed by a recovery of the hardening trend. The M(3) mode shows a moderate slope change near 12.6 GPa and disappears at 17.3 GPa, where a new mode $\omega$(1) emerges at 82 cm$^{-1}$. This new mode hardens with pressure and exhibits a pronounced slope change around 20.3 GPa. Such multiple slope changes, mode mergings, and the appearance of new Raman features correlate strongly with the sequence of pressure-driven isostructural distortions and the eventual onset of the long-range structural transition, indicating their vibrational sensitivity to subtle symmetry modifications. 
	
		\begin{figure*}[ht!]
\includegraphics[width=18cm]{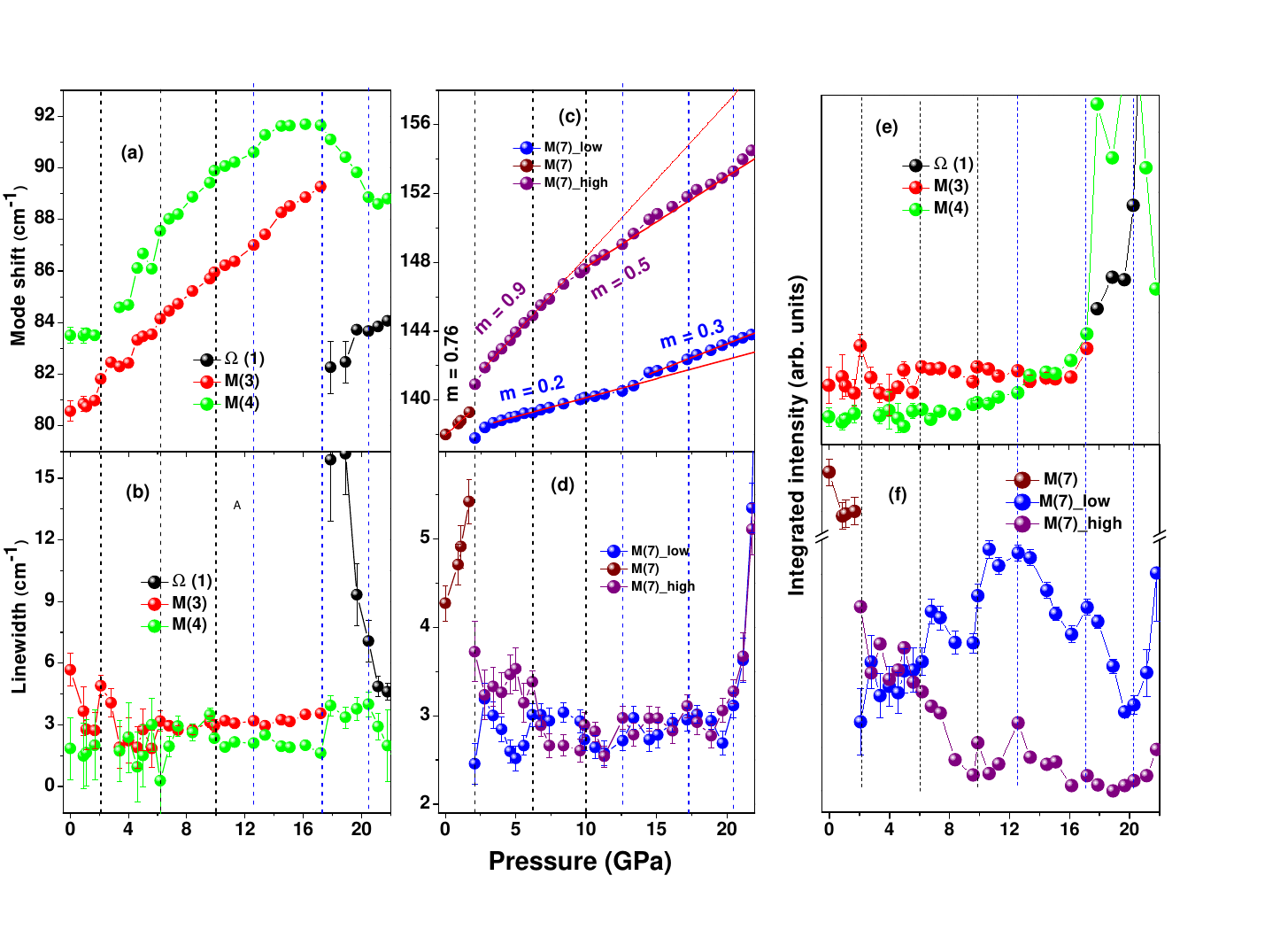}
\caption{\label{fig6} Pressure evolution of the vibrational properties of low frequency Raman modes. (a)–(c) Mode frequency; (b)–(d) corresponding mode linewidths; and (e)–(f) integrated mode intensities. Vertical indicators denote structural milestones: black dashed lines mark isostructural transitions, while blue dashed lines highlight the onset of long-range symmetry transitions or major structural rearrangements.} 
\end{figure*}
		
	The linewidth evolution follows a similar sequence: the M(4) mode remains nearly constant up to ~2 GPa, whereas the M(3) mode gradually narrows in this range (Fig. 6(b)). At 2.1 GPa, the linewidth of the M(4) mode broadens significantly due to mode merging. At higher pressures (2.1–17.3 GPa), the M(3) mode exhibits only minor linewidth variations. In contrast, the M(4) mode shows moderate irregularities around 6.2 and 9.9 GPa. Above 17.3 GPa, the M(4) mode broadens sharply and displays a plateau-like behavior between 18 and 20.3 GPa, before narrowing again at higher pressures. The newly emerged $\omega$(1) mode, on the other hand, shows rapid linewidth narrowing with continued compression.		
				
		Figures 6(c) and 6(d) present the pressure evolution of the mode frequency and linewidth of the M(7) mode. This mode initially hardens up to 2 GPa with a rate of 0.76 cm$^{-1}$/GPa. In contrast to the typical linewidth narrowing expected under compression, its linewidth steadily broadens in this range, indicating anomalous scattering behavior. Such broadening suggests the onset of lattice instabilities or the activation of lattice–orbital interactions. Consequently, at 2.1 GPa, the mode splits into two distinct components at 137.8 (M(7)$_{low}$) and 140.9 cm$^{-1}$ (M(7)$_{high}$), signaling local symmetry breaking (Figs. 6(c)). With increasing pressure, the two components exhibit markedly different responses. The high frequency branch M(7)$_{high}$ mode hardens at a higher rate (0.9 cm$^{-1}$/GPa), whereas the lower branch M(7)$_{low}$ mode shows only weak hardening (0.2 cm$^{-1}$/GPa). The M(7)$_{high}$ mode displays discernible slope changes at around  9.9, and 20.5 GPa, while the M(7)$_{low}$ mode shows a distinct slope anomaly at 12.6 GPa. Their linewidths also evolve in opposite ways between 2.4 and 10 GPa, with a crossover occurring near 6.2 GPa. Above 10 GPa, both modes exhibit comparable linewidth evolution, albeit with moderate irregularities around 12.6 and 17.3 GPa. Beyond 20.5 GPa, both modes undergo substantial linewidth broadening and converge to similar linewidth values.

The integrated intensity (spectral weight) of the M(4) mode remains nearly unchanged up to 6 GPa, after which it gradually increases up to 17.3 GPa, exhibiting a clear anomaly near 12.6 GPa—coinciding with the onset of the long-range structural transition (Fig. 6(e)). In contrast, the neighboring M(3) mode shows a pronounced maximum at 2.1 GPa, and its intensity becomes significantly higher than that of the M(4) mode between 6.2 and 12.6 GPa. At 13.4 GPa, the intensities of the two modes become comparable up to 17.3 GPa, beyond which the M(3) mode disappears. Above 17.3 GPa, the integrated intensities of both M(4) mode and newly emerged $\omega$(1) mode increase sharply; however, the intensity of the M(4) mode begins to drop rapidly beyond 20.3 GPa. The integrated intensity evolution of the M(7) mode and its two split branches also reveals highly intriguing behavior (Fig. 6(f)). The intensity of the parent mode decreases with pressure up to 2 GPa, before it splits into two components at 2.1 GPa. At the splitting point, the higher-frequency branch exhibits greater intensity, but the two branches display comparable intensities between 2.4 and 6.2 GPa. Beyond this pressure, the higher-frequency branch shows a rapid decrease in intensity up to 9.9 GPa, whereas the lower-frequency branch gains intensity strongly up to 12.6 GPa, with a distinct irregularity near 9.9 GPa. Around 12.6 GPa, the higher-frequency branch exhibits a peak-like feature, followed by a continuous reduction up to 17.3 GPa and only marginal changes between 17.3 and 20.3 GPa. In contrast, the lower-frequency branch gradually loses intensity from 12.6 to 20.5 GPa, showing a discontinuity around 17.3 GPa, and then begins to increase again above 20.3 GPa. Notably, the overall integrated intensity of the lower-frequency branch increases dramatically in the 2.1–12.6 GPa range. In the absence of any long-range symmetry change, such pronounced spectral-weight redistribution and mode-specific anomalies likely indicate the activation of additional scattering channels—most plausibly involving orbital–lattice interactions and/or spin–phonon coupling.
	
			In Fig. 7(a), the pressure evolution of mode frequency of three closely spaced Raman modes M(8), M(9), and M(10) are shown, with their corresponding linewidths presented in Fig. 7(b). All three modes exhibit gradual hardening with increasing pressure; however, their pressure coefficients differ markedly. The M(8) and M(10) modes harden at a rate of ~0.6 cm$^{-1}$/GPa, whereas the intermediate M(9) mode hardens at nearly twice this rate. All modes begin to deviate from these linear trends around 6.2 GPa, with the deviations becoming more prominent above 9.9 GPa. The linewidths of all three modes display typical narrowing up to 2.1 GPa (Fig. 7(b)). Between 2.1 and 6.2 GPa, the linewidths of the M(8) and M(9) modes remain nearly constant, followed by a finite drop and then gradual broadening up to 10.5 GPa. Additional anomalies are visible in their linewidth evolution around 17.3 and 20.5 GPa. In contrast, the M(10) mode exhibits pronounced broadening between 2.1 and 10.5 GPa, with a clear irregularity near 6.2 GPa. The nearly twofold increase in linewidth is particularly noteworthy and cannot be attributed solely to local symmetry breaking or the isostructural transition at 2.1 GPa. Above 10.5 GPa, this mode undergoes rapid narrowing and eventually disappears above ~18 GPa. Such pronounced and mode-selective linewidth anomalies strongly suggest the activation of additional scattering channels—possibly involving spin–phonon and/or orbital–lattice coupling under pressure—rather than being solely a consequence of structural distortions.

\begin{figure*}[ht!]
\includegraphics[width=18cm]{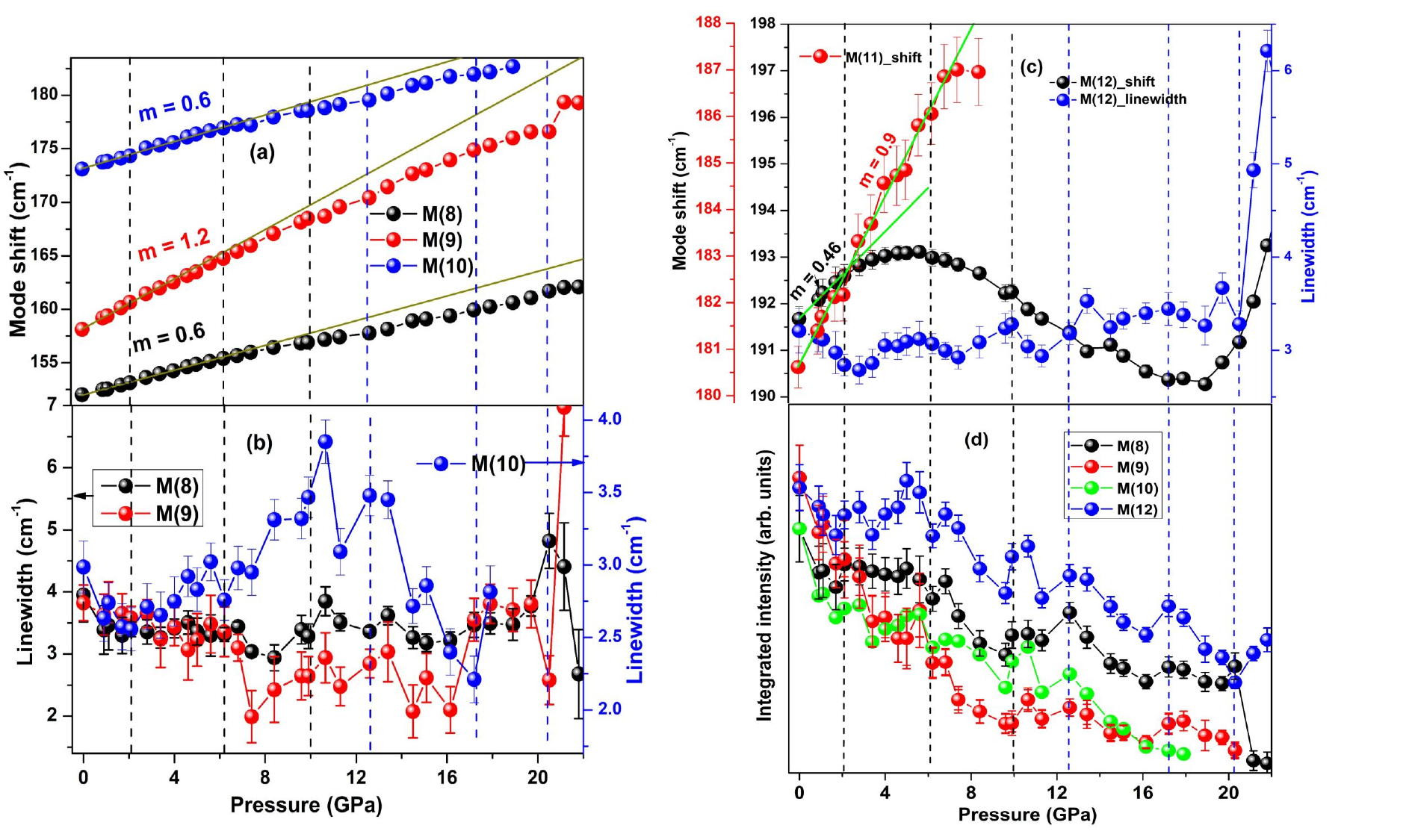}
\caption{\label{fig7} Pressure dependence of (a) mode frequencies and (b) linewidths for modes M(8)-M(10); (c) frequency evolution of M(11) alongside the frequency shift and linewidth variation of M(12); (d) integrated intensities for modes M(8)-M(10) and M(11). Vertical black dashed lines denote isostructural transitions, while blue dashed lines highlight long-range symmetry transitions or major structural rearrangements.} 
\end{figure*}

A particularly important behavior is observed in the pressure evolution of the 191.5 cm$^{-1}$ (M(12)) mode (Fig. 7(c)). This mode hardens with a slope of 0.46 cm$^{-1}$/GPa up to 2.1 GPa, beyond which it gradually deviates from the initial linear trend, producing a concave-like curve. Notably, the slope of this mode is nearly half that of the neighboring M(11) mode (m = 0.9 cm$^{-1}$/GPa), suggesting that the short-span linear trend of the M(12) mode is actually part of a broader softening behavior. Around 6 GPa, it begins to soften in absolute terms and continues up to 17.3 GPa, where its frequency becomes nearly constant in the 17.3–19 GPa range, followed by a rapid hardening above 20.5 GPa. The frequency–pressure curve of this mode also shows a distinct anomaly at 12.6 GPa and a slight irregularity near 9.9 GPa. Linewidth analysis further reveals pronounced peculiarities, exhibiting either minima or maxima around 2.1, 6, 9.9, 12.6, 17.3, and 20.5 GPa. Taken together, the anomalous frequency and linewidth evolution of the M(12) mode strongly suggest that its softening plays a decisive role in governing potential orbital–lattice–spin correlations and structural instabilities, from local symmetry breaking to the onset of long-range symmetry transition or major structural rearrangements. In this sense, the M(12) mode acts as a fingerprint for tracking the onset and progression of the structural phase transformation in NNO.

To further elucidate the origin and extent of these instabilities, we examine the pressure evolution of the integrated intensities of the M(12) along with  M(8)-M(10) modes. The integrated intensity of the M(12) mode evolves in a notably complex fashion. It gradually decreases up to 2.1 GPa, followed by a clear enhancement between 2.1 and 6.2 GPa, and then shows a continuous reduction up to 20.5 GPa, accompanied by distinct irregularities around 9.9, 12.6, and 17.3 GPa (Fig. 7(d)). Above 20.5 GPa, this mode again begins to gain intensity, consistent with a major structural rearrangement at high pressure. The integrated intensities of the other three modes M(8)-M(10) exhibit an overall gradual reduction in the 0–2.1 GPa range. Beyond this region, the M(8) and M(9) modes show a general decreasing trend with pronounced discontinuities near 6.2, 9.9, and 17.3 GPa. In contrast, the M(10) mode remains nearly unchanged between 2.1 and 6.2 GPa, above which decreases steadily up to 10 GPa, and then shows a marginal increase up to 12.6 GPa. Between 12.6 and 20.5 GPa, its intensity again decreases with a clear anomaly near 17.3 GPa, and above 20.5 GPa it drops sharply.

In Fig. 8, the pressure evolution of the M(6) and M(14) modes and their subsequent splitting into two distinct branches at 6.2 and 9.9 GPa are presented. The M(6) mode exhibits marginal hardening up to 6.2 GPa, with a slight reduction in slope from 0.34 cm$^{-1}$/GPa in 0–2.1 GPa to 0.24 cm$^{-1}$/GPa in 2.1–6.2 GPa (Fig. 8(a)). At 6.2 GPa, it splits into two distinct components at 129.1 (M(6)$_{low}$) and 132.4 cm$^{-1}$ (M(6)$_{high}$). The linewidth of the parent mode gradually narrows up to 2.1 GPa, followed by progressive broadening until the onset of splitting at 6.2 GPa as shown in Fig. 8(b). Notably, despite the subtle anomaly in the mode shift, the linewidth exhibits a sharp minimum, indicating the involvement of additional scattering processes beyond simple structural modifications. After splitting, the higher-frequency branch M(6)$_{high}$ shows slight hardening at a rate of 0.3 cm$^{-1}$/GPa up to 9.9 GPa, followed by gradual softening and eventual disappearance above 20.5 GPa. In contrast, the lower-frequency branch M(6)$_{low}$ undergoes continuous softening, initially at a rate of –0.8 cm$^{-1}$/GPa in 6.2–12.6 GPa, followed by a faster softening in 12.6–20.5 GPa, resulting in a curved-like pressure evolution. Beyond 20.5 GPa, it exhibits linear softening. The linewidths of both branches remain nearly constant between 6 and 13.6 GPa. Beyond this range, the lower-frequency branch broadens progressively up to 20.5 GPa with an irregularity around 17.3 GPa, then narrows at higher pressures. The higher-frequency branch, on the other hand, undergoes a rapid linewidth decrease, becoming too weak to extract beyond 15.3 GPa, although faint traces persist up to 20.5 GPa.

\begin{figure*}[ht!]
\includegraphics[width=18cm]{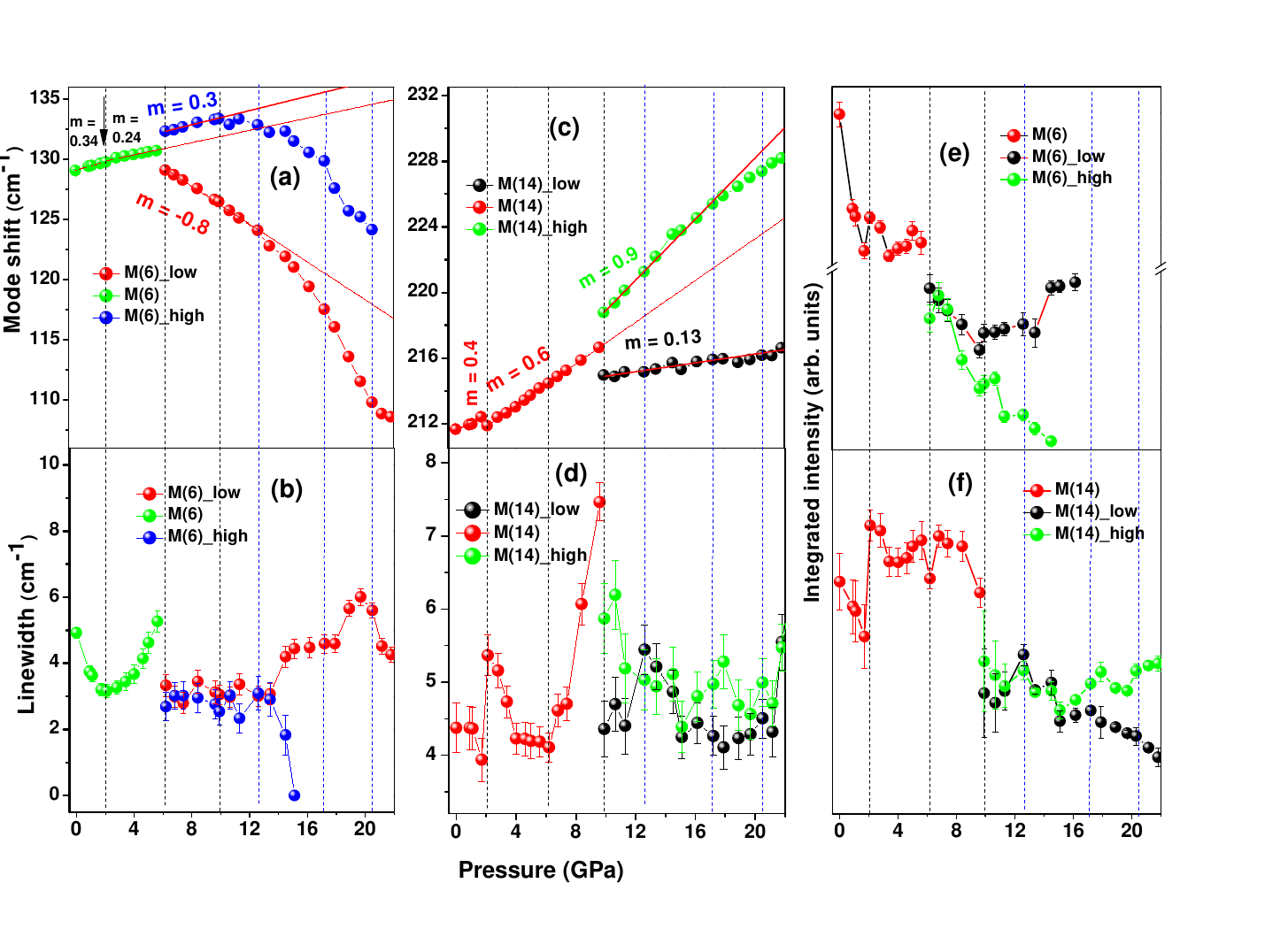}
\caption{\label{fig8} Pressure-dependent Raman evolution of the M(6) and M(14) modes. (a) Frequency shift and (b) linewidth for the M(6) mode and its low- and high-frequency branches (M(6)$_{low}$ and M(6)$_{high}$ ); (c) frequency shift and (d) linewidth for the M(14) mode and its corresponding branches M(14)$_{low}$ and M(14)$_{high}$); and integrated intensity evolution for (e) M(6) and (f) M(14) along with their respective branches. Vertical black dashed lines denote isostructural transitions, while blue dashed lines highlight long-range symmetry transitions or major structural rearrangements.} 
\end{figure*}

	The M(14) mode exhibits a subtle discontinuity and slope change at 2.1 GPa during its hardening up to 9.9 GPa, where it subsequently splits into two branches at 215 (M(14)$_{low}$) and 218.6 cm$^{-1}$  M(14)$_{high}$ as manifested in Fig. 8(c). The linewidth of the M(14) mode remains nearly constant initially, with a slight drop near 2 GPa, followed by pronounced broadening at 2.1 GPa, and then narrows up to 6.2 GPa (Fig. 8(d)). In the 6.2–9.9 GPa range, the linewidth broadens significantly, eventually leading to the mode splitting. Post-splitting, the lower-frequency branch M(14)$_{low}$ shows only marginal hardening at a rate of only 0.13 cm$^{-1}$/GPa, whereas the higher-frequency branch M(14)$_{high}$ hardens more rapidly with the rate 0.9 cm$^{-1}$/GPa. Both branches exhibit linewidth anomalies near 12.6 and 17.3 GPa, followed by drastic broadening trend above 20.5 GPa. 

	The integrated intensity of the M(6) mode also shows a characteristic multi-stage evolution under pressure (Fig. 8(e)). Its intensity decreases initially and remains nearly unchanged in the 2.1–6.2 GPa range, consistent with the  anomaly already observed in its linewidth behavior. After the mode splits at 6.2 GPa, both branches exhibit intensity reduction in the 6.2-9.9 GPa interval, with the higher-frequency branch decaying more rapidly. Beyond 9.9 GPa, the lower-frequency branch M(6)$_{low}$ maintains nearly constant intensity up to 17.3 GPa before showing a gradual enhancement at higher pressures, whereas the higher-frequency branch M(6)$_{high}$ continues to weaken and becomes increasingly suppressed.
 
	A similar yet more pronounced trend is observed in the M(14) mode. Its integrated intensity initially decreases up to 2 GPa, mirroring the behavior of the M(6) mode, followed by a sharp increase around 2.1 GPa and another distinct irregularity near 6.2 GPa, both consistent with the onset of pressure-induced instabilities detected in several other modes (Fig. 8(f)). Above 6.2 GPa, the intensity progressively declines until the mode splits at 9.9 GPa. Post-splitting, the two branches exhibit comparable intensities and display a clear anomaly at 12.6 GPa. Above 17.3 GPa, their intensities start to diverge: the lower-frequency branch decreases gradually, while the higher-frequency branch shows an overall increase, indicating distinct coupling strengths and mode-selective participation in the high-pressure structural evolution.
	
	In Fig. S3 \cite{SM}, the pressure evolution of the mode frequency, linewidth, and integrated intensity for the M(18) (270 cm$^{-1}$), and two high-frequency modes at 691 (M(39)) and 833 cm$^{-1}$ (M(40)) is presented. The  M(18) mode hardens linearly up to 6.2 GPa with a rate of 1.8 cm$^{-1}$/GPa, but begins to soften relative to this trend beyond this pressure (Fig. S3(a)). Its linewidth narrows up to 5 GPa and remains nearly unchanged in the 5–7 GPa range, indicating an additional scattering channel active near 6.2 GPa. Around 9.9 GPa, the mode abruptly narrows, followed by distinct anomalies at 12.6, 17.3, and 20.5 GPa. In contrast, the M(39) and M(40) modes exhibit a monotonic blue shift up to 9.9 GPa (Figs. S3(b) and S3(c)), albeit with different rates (4.1 and 2.5 cm$^{-1}$/GPa, respectively). Above 9.9 GPa, the M(40) mode displays a gradual softening relative to its initial hardening trend, while the M(39) mode shows only a subtle deviation above 12.6 GPa. The high hardening rate of the M(39) mode, together with its minimal sensitivity to the isostructural transitions and to possible orbital–lattice–spin coupling, suggests that this mode frequency is only affected by the structural rearrangements. Nevertheless, its linewidth shows weak anomalies near 2.1 and 10 GPa and more pronounced renormalization at 12.6 and 17.3 GPa. The linewidth of the M(40) mode steadily narrows up to 17.3 GPa, above which it starts to broaden, producing a minimum at the pressure 17.3 GPa, where its frequency begins to redshift.

	Interestingly, the integrated intensity of the M(39) mode undergoes strong renormalization at high pressures, despite its frequency being least perturbed (Fig. S3(d)). The intensity decreases up to about 2 GPa, followed by an abrupt enhancement at 2.1 GPa and only marginal variation up to 6.2 GPa; it then decreases again up to 9.9 GPa. Above 12.6 GPa, the intensity increases noticeably and then rapidly decreases beyond 15 GPa, with a clear discontinuity around 17.3 GPa. The other two modes M(18) and M(40) show weaker irregularities around 2.1 and 6.2 GPa, but display pronounced renormalization near 9.9 GPa (Figs. S3(d) and S3(e) ). At higher pressures, the M(18) mode exhibits strong anomalies at 12.6, 17.3, and 20.3 GPa, while the M(40) mode shows moderate discontinuities at these pressures.

	\subsection{High pressure synchrotron XRD study}
	
	High-pressure synchrotron radiation XRD patterns of NNO in the pressure range 4.3–34 GPa are presented in Fig. 9. Up to 10.6 GPa, all diffraction patterns exhibit a uniform shift of Bragg peaks toward higher diffraction angles, consistent with normal lattice contraction, without any additional noticeable features. In this pressure range, the XRD patterns are satisfactorily fitted to the ambient \textit{Pbcn} structural model, and a representative Rietveld refinement at 9 GPa is shown in Fig. 10(a). At 12.7 GPa, splitting of a Bragg reflection is observed around 2$\theta$ = 8.23$^{\circ}$, accompanied by the emergence of three additional peaks near 12.27, 14.52, and 21.14$^{\circ}$, which gradually intensify with pressure. This marks the onset of a long-range structural transition, consistent with the appearance of the octahedral mode at 12.6 GPa. At higher pressures, additional Bragg peaks emerge at 21.6 GPa, notably at 13.5$^{\circ}$ and 19.92$^{\circ}$ (purple arrows in Fig. 9), while several reflections develop asymmetric broadening, suggesting tailing features associated with new diffraction lines. Further new Bragg reflections are detected at 26.6 GPa (highlighted by orange arrows), indicating a significant transformation of the ambient structure into the high-pressure phase. Indexing of these new reflections at 26.6 GPa yields the best match with the monoclinic \textit{P2/c} phase, a direct subgroup of the ambient \textit{Pbcn} symmetry, with lattice parameters comparable to its Mn analogue. Rietveld refinement confirms that the XRD patterns throughout the 12.3–34 GPa range are well described by a coexistence of \textit{Pbcn} and \textit{P2/c} phases. Representative refinements at 26.6 and 34 GPa are shown in Figs. 10(b)-(c). Illustrations of the unit cells for the \textit{P2/c} structures at the highest pressure of 34 GPa are provided in Fig. S4, while the corresponding refined lattice parameters and structural coordinates are listed in Table S II \cite{SM}.

\begin{figure*}[ht!]
\includegraphics[width=14cm]{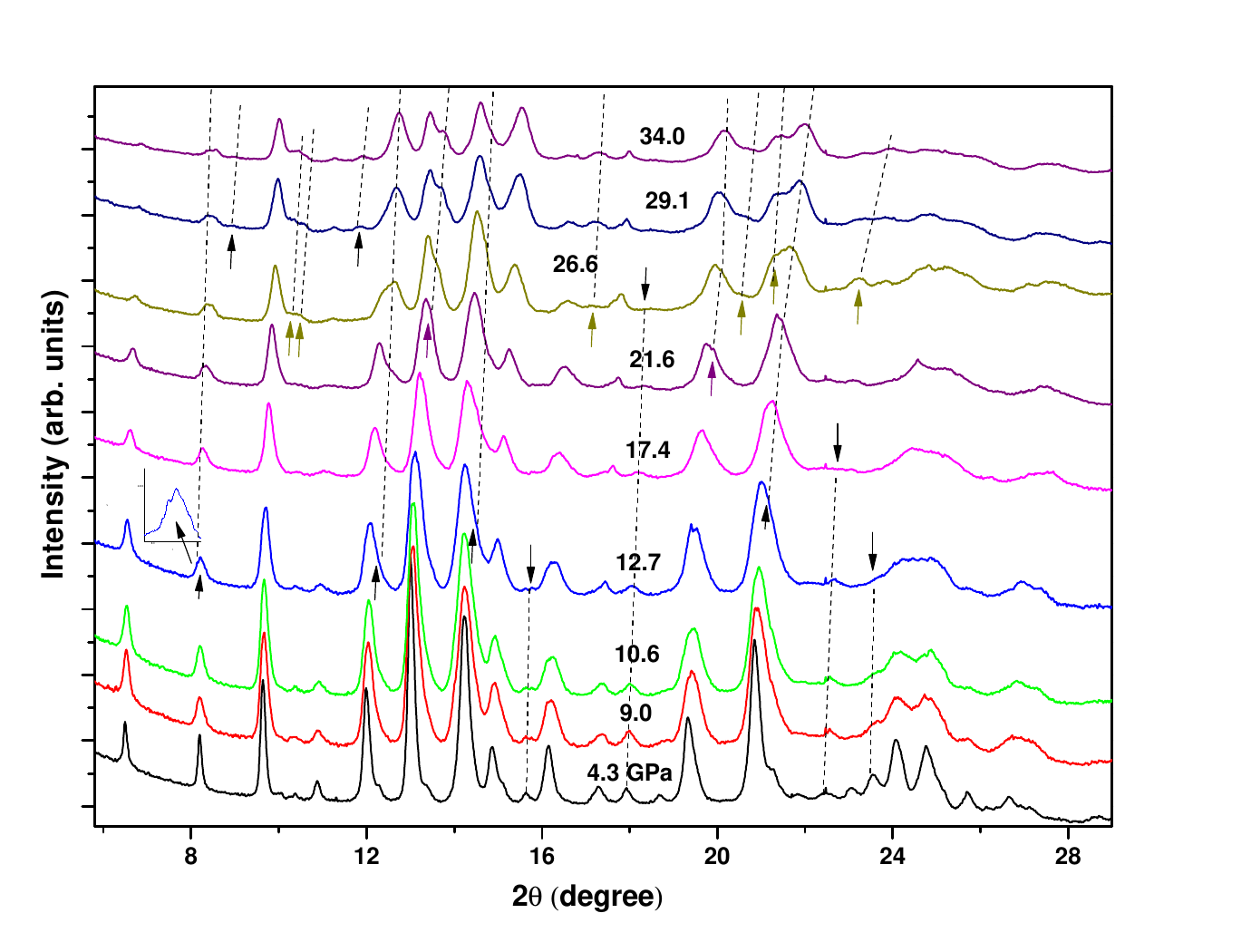}
\caption{\label{fig5} High pressure XRD patterns of NNO up to 34 GPa. The emergence of new diffraction peaks is highlighted by upward arrows, while those that disappear are marked by downward arrows; the evolution of these peaks is traced by black dashed lines.} 
\end{figure*}

\begin{figure*}[ht!]
\includegraphics[width=12cm]{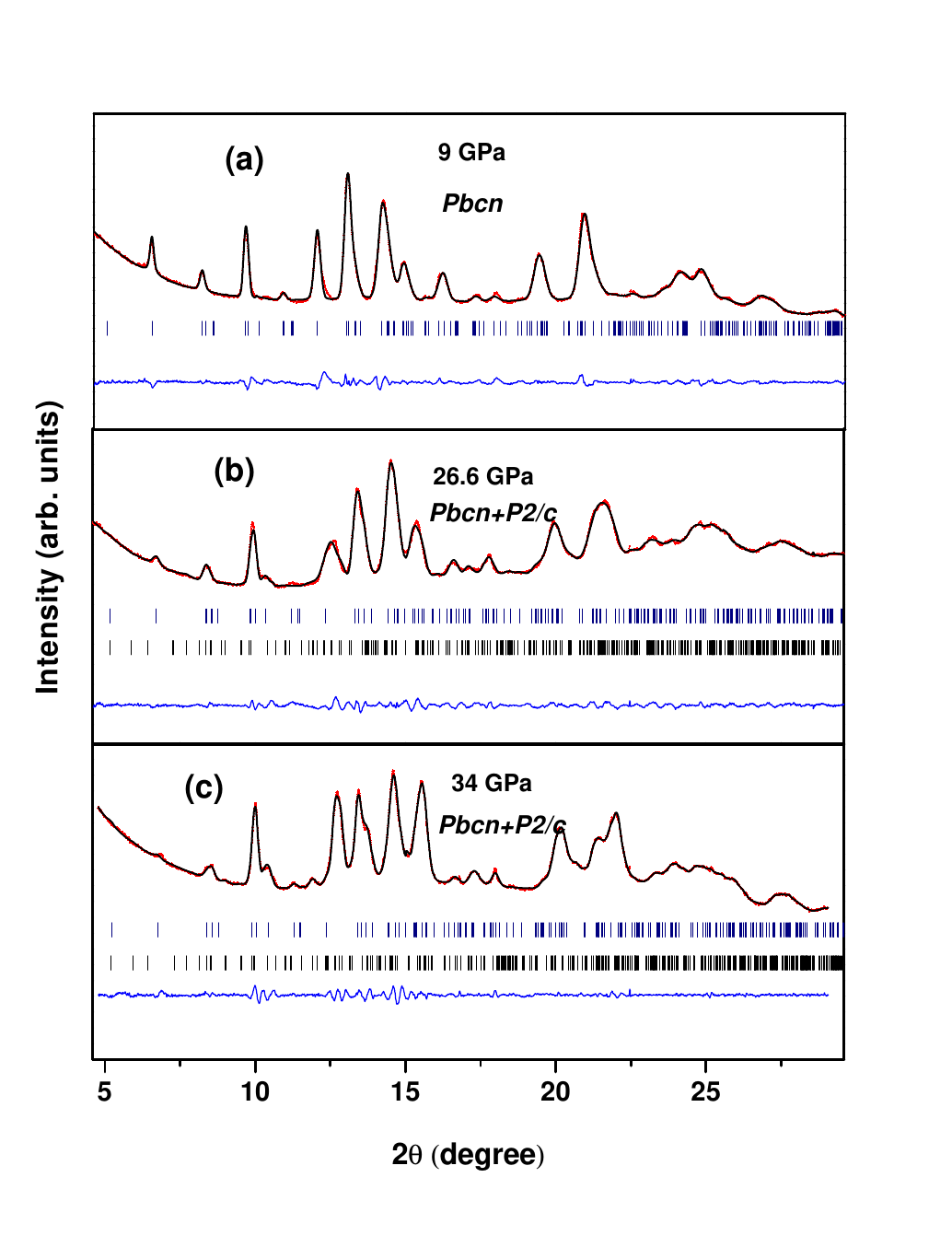}
\caption{\label{fig10} Rietveld refined XRD patterns at (a) 9 GPa, (b) 26.6 GPa, (c) 34 GPa. The solid red circles represent the experimental data points, and the black lines correspond to the structural fits. Allowed Bragg positions for the \textit{Pbcn} and \textit{P2/c} phases are identified by navy and black vertical ticks, respectively, while the blue line at the bottom represents the difference between the experimental data and the theoretical fit.} 
\end{figure*}

	Fig. 11(a) shows the pressure evolution of the normalized lattice parameters (\textit{L/L$_0$}) in the \textit{Pbcn} phase. A clear slope change in all three lattice parameters appears near 6 GPa, with the $b$ axis exhibiting noticeably smaller compression compared to $a$ and $c$. This anomaly is fully consistent with the Raman irregularity observed at 6.2 GPa. Due to insufficient XRD data points near 2 GPa, the earlier Raman anomaly at ~2.1 GPa cannot be independently investigated. Above 6 GPa, the $a$ parameter continues to contract progressively up to 21 GPa. In sharp contrast, both $b$ and $c$ parameters show only marginal compression in the 6–10 GPa range. Beyond 10 GPa, $c$ parameter exhibits a slight expansion until ~12.7 GPa, followed by gradual compression up to 21 GPa. At higher pressures, $a$ and $b$ show only minimal contraction, whereas the $c$ axis compresses much more rapidly. 
	
\begin{figure*}[ht!]
\includegraphics[width=14cm]{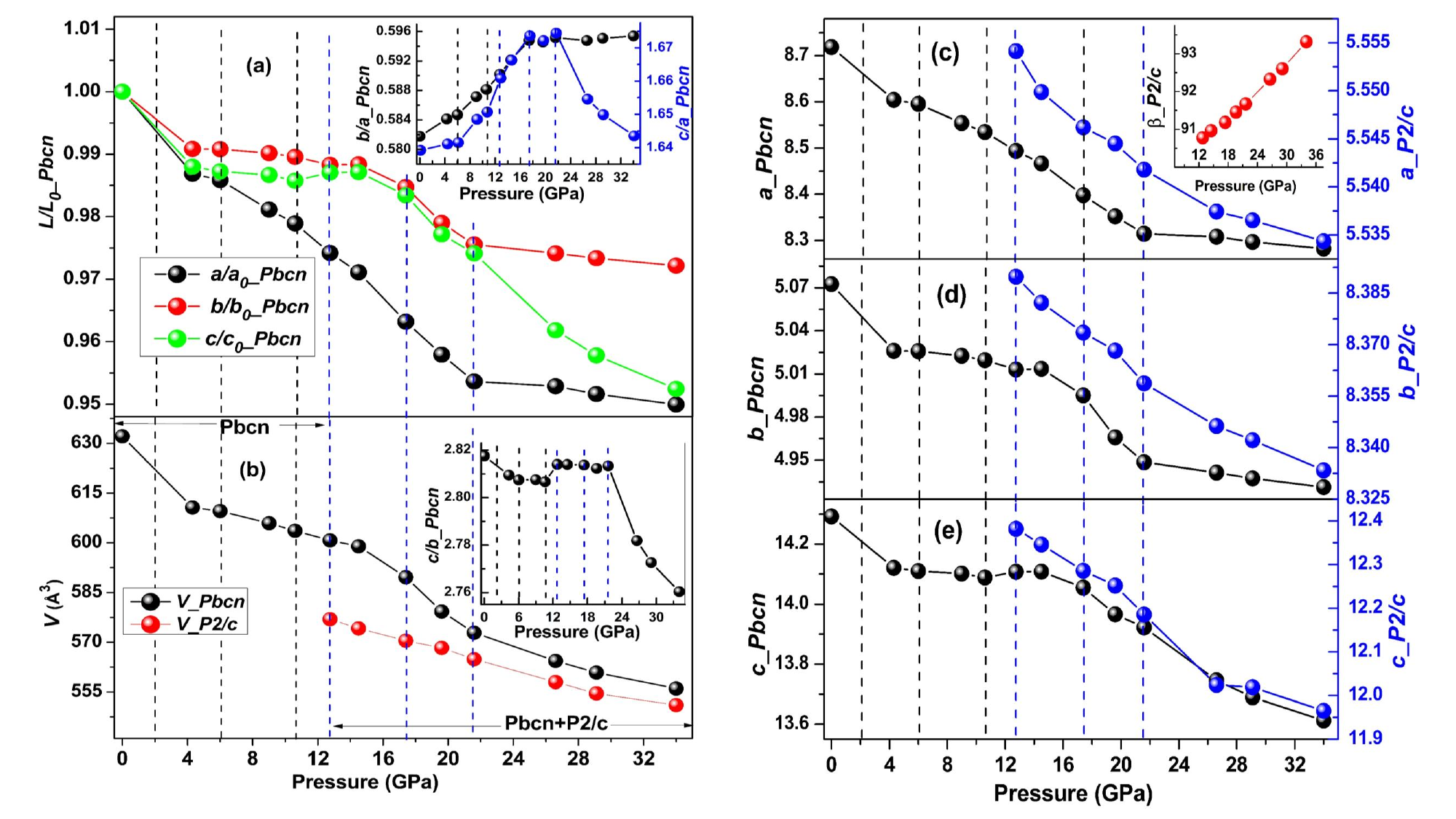}
\caption{\label{fig11} Pressure evolution of (a) normalized lattice parameters (\textit{L/L$_0$}) for the \textit{Pbcn} phase, and (b)  unit cell volume (\textit{V} ) for the \textit{Pbcn} and \textit{P2/c} phases. Pressure dependence of absolute lattice parameters (c) \textit{a}, (d) \textit{b}, (e) \textit{c} for both \textit{Pbcn} and  \textit{P2/c} phases. The inset in (d) displays the pressure evolution of the monoclinic angle ($\beta$) for the \textit{P2/c} phase.} 
\end{figure*}
	
	This highly anisotropic compression leads to multiple anomalies in the axial ratios. In the 0–12.7 GPa range, the strongest compression occurs along $a$, causing both $b/a$ and $c/a$ ratios to increase with pressure. A distinct slope change in the $c/a$ ratio at 6 GPa is visible in the inset of Fig. 11(a). Around 17.3 GPa, both $b/a$ and $c/a$ ratios level off and remain nearly constant up to 21 GPa, indicating nearly isotropic compression in this interval. Above 21 GPa, the $c/a$ ratio decreases rapidly due to the enhanced compressibility of the $c$ axis. The $c/b$ ratio, shown in Fig. 11(b) inset, decreases steadily up to 10 GPa with a clear slope change near 6 GPa, followed by a significant increase in 10–12.7 GPa and an almost constant trend from 12.7–21 GPa; above this, it drops sharply, mirroring the $c/a$ behavior. Overall, the XRD anomalies closely track those identified in the Raman study.

	Figures 11(c)–(e) present the pressure dependence of absolute lattice parameters for both the ambient \textit{Pbcn} and high-pressure \textit{P2/c} phases. The \textit{P2/c} lattice parameters compress more systematically than those of \textit{Pbcn}, although subtle irregularities are evident around 17.3 and 21 GPa. An additional slope change is observed in the $c$ parameter at 26.6 GPa. The monoclinic angle of the \textit{P2/c} phase increases steadily with pressure (inset of Fig. 11(c)). The unit-cell volume in the \textit{P2/c} phase contracts smoothly from 12.7 to 34 GPa, with a moderate slope change around 21 GPa (Fig. 11(b)). In contrast, the \textit{Pbcn} phase shows slower compression in 6–12.7 GPa compared to 0–4.3 GPa, followed by a higher compression rate in the 14–21 GPa range. Remarkably, above 21 GPa, both Pbcn and \textit{P2/c} phases exhibit nearly identical volume-compression rates.


\subsection{Discussion}

	The combined Raman and and synchrotron XRD  establish a coherent picture of a pressure-driven vibrational, structural and vibrational evolution of NNO, governed by a hierarchy of lattice instabilities coupled to potential orbital and magnetic degrees of freedom. First, the Raman spectra reveal a sequence of anomalies at ~2.1, 6.2, 10, 12.6–13.4, 17.3, and 20.5 GPa across multiple phonon modes, manifested through mode softening, slope changes, linewidth anomalies, splitting, and emergent modes. These signatures provide compelling spectroscopic evidence of pressure-induced modifications in local bonding geometry and symmetry. Among them, the 191.5 cm$^{-1}$ (M(12)) mode stands out as a key instability marker, displaying a transition from normal hardening to pronounced softening and multiple characteristic irregularities in both mode frequency and linewidth. Between 2 and 12.6 GPa, the combination of anomalous linewidths, anisotropic mode shifts ($+1.2$ to $-0.8$ cm$^{-1}$/GPa) in the low-frequency regime, and the strengthening 137 cm$^{-1}$ low-frequency branch, is incompatible with purely anharmonic lattice compression. Instead, these features point to the activation of strong orbital–lattice and/or spin–phonon interactions, mirroring the reported low-temperature behavior \cite{Singh, Jana6}. The emergence of new Raman modes and branch splitting at several characteristic pressures further confirms pressure-induced lowering of local symmetry and dynamic distortions. 
	
	The synchrotron XRD results independently corroborate these spectroscopic anomalies. Distinct slope changes in lattice parameters and their ratios at ~6 and 10 GPa closely correspond to the Raman anomalies at 6.2 and 10 GPa, suggesting the (isotructural) onset of incipient structural rearrangements within the \textit{Pbcn} phase. Notably, previous low-temperature study reported magneto-lattice coupling associated with an enhancement of the $b/a$ ratio near the ferrimagnetic ordering temperature \cite{Tailleur}, while composition-dependent studies \cite{Bolletta, Tarakina, Jiongo-Dongmo, Martin2} have linked a decreasing $b/a$ ratio to a reduction in ordering temperature. Consequently, the observed pressure-induced increase in the $b/a$ ratio could potentially modify the magnetic exchange interactions, offering a possible route to tune the magnetic ordering temperature. The XRD data also reveal that above ~14 GPa, compression becomes increasingly anisotropic before evolving toward an isotropic regime near 17–21 GPa. Crucially, both Raman and XRD converge on the onset of a major structural transformation around 17.3 and 20.5–21 GPa, where rapid mode hardening/softening, linewidth broadening, and intensity redistribution accompany marked changes in lattice parameters and volume compression. This transition corresponds to the emergence and gradual stabilization of the high-pressure monoclinic \textit{P2/c} phase. The increasing monoclinic angle and systematic compression of the \textit{P2/c} lattice parameters emphasize the robustness of this phase above ~21 GPa. 
	
	Taken together, these results indicate that pressure drives a multi-stage structural evolution—from local symmetry breaking and isostructural distortions in the \textit{Pbcn} phase (2–12.7 GPa), through potential orbital-lattice-spin correlations and mode softening (14–20 GPa), ultimately culminating in a long-range symmetry transformation to the \textit{P2/c} phase. Remarkably, the trigonal Mn analogue, Mn$_4$Nb$_2$O$_9$, exhibits an almost identical pressure response, displaying three isostructural transitions near 2, 6.2, and 10 GPa, closely mirroring those observed in NNO \cite{Jana5}. Furthermore, both systems show the onset of partial transformation to the $P2/c$ phase around 13 GPa, followed by a pronounced conversion from the ambient phase to the high-pressure $P2/c$ phase in the 22–23 GPa range, as evidenced by the drastic reorganization of their Raman spectra. In contrast, Co- and Fe-based analogues (CNO and FNO) undergo qualitatively different long-range structural transitions at substantially different pressures \cite{Jana2, Jana3, Sahu}. The striking similarities between NNO and MNO thus reflect their close local structural correspondence, as evidenced by similar NMR parameters, including quadrupolar frequency and nuclear relaxation constants \cite{Jana4}. Overall, the combined Raman and XRD results suggest the intricate coupling among lattice, orbital, and magnetic degrees of freedom in NNO, underscoring the pivotal role of local structural environment in governing the high-pressure behavior of honeycomb-layered magnetoelectrics.

\section{Conclusion}
In summary, comprehensive NMR, Raman, and synchrotron XRD measurements reveal that Ni$_4$Nb$_2$O$_9$ undergoes a highly pressure-sensitive, multistage structural evolution governed by potential lattice–orbital–spin coupling. Despite its orthorhombic average symmetry, ambient-pressure NMR establishes a close local structural similarity between and its trigonal analogue Mn$_4$Nb$_2$O$_9$, which underpins their remarkably similar pressure-induced transition sequences. With increasing pressure, successive isostructural distortions precede an incipient symmetry-lowering transition from \textit{Pbcn} to $P2/c$, ultimately leading to a stabilized high-pressure phase. The anomalous softening and renormalization of Raman phonons—particularly the 191.5  cm$^{-1}$ (M(12)) mode—provide direct evidence of emerging structural instabilities and potential spin–orbital–lattice correlations. These findings underscore the pivotal role of local structure in dictating the high-pressure behavior of honeycomb-layered materials and offer key insights into tuning functional properties in correlated magnetic systems.

\section{Acknowledgments}
	This work is supported by the National Science Foundation of China (W2432007, 42150101), the National Key Research and Development Program of China (2022YFA1402301), Shanghai Key Laboratory for Novel Extreme Condition Materials, China (No. 22dz2260800), Shanghai Science and Technology Committee, China (no. 22JC1410300). Raman scattering data are collected using Renishaw inVia system at the University of Texas at Austin. The authors gratefully acknowledge SPring-8 for providing synchrotron radiation facilities. Experiments at the BL12B2 sector of SPring-8 are conducted under Proposal No. 2025A4141. T.N. gratefully acknowledges funding from the Beijing Natural Science Foundation (Project No. IS24025).

\newpage

\section{Supplemental material}     

\setcounter{figure}{0}
\renewcommand{\figurename}{Fig.}
\renewcommand{\thefigure}{S\arabic{figure}}
\renewcommand{\tablename}{Table S}

\begin{figure}[ht!]
\includegraphics[width = 12cm]{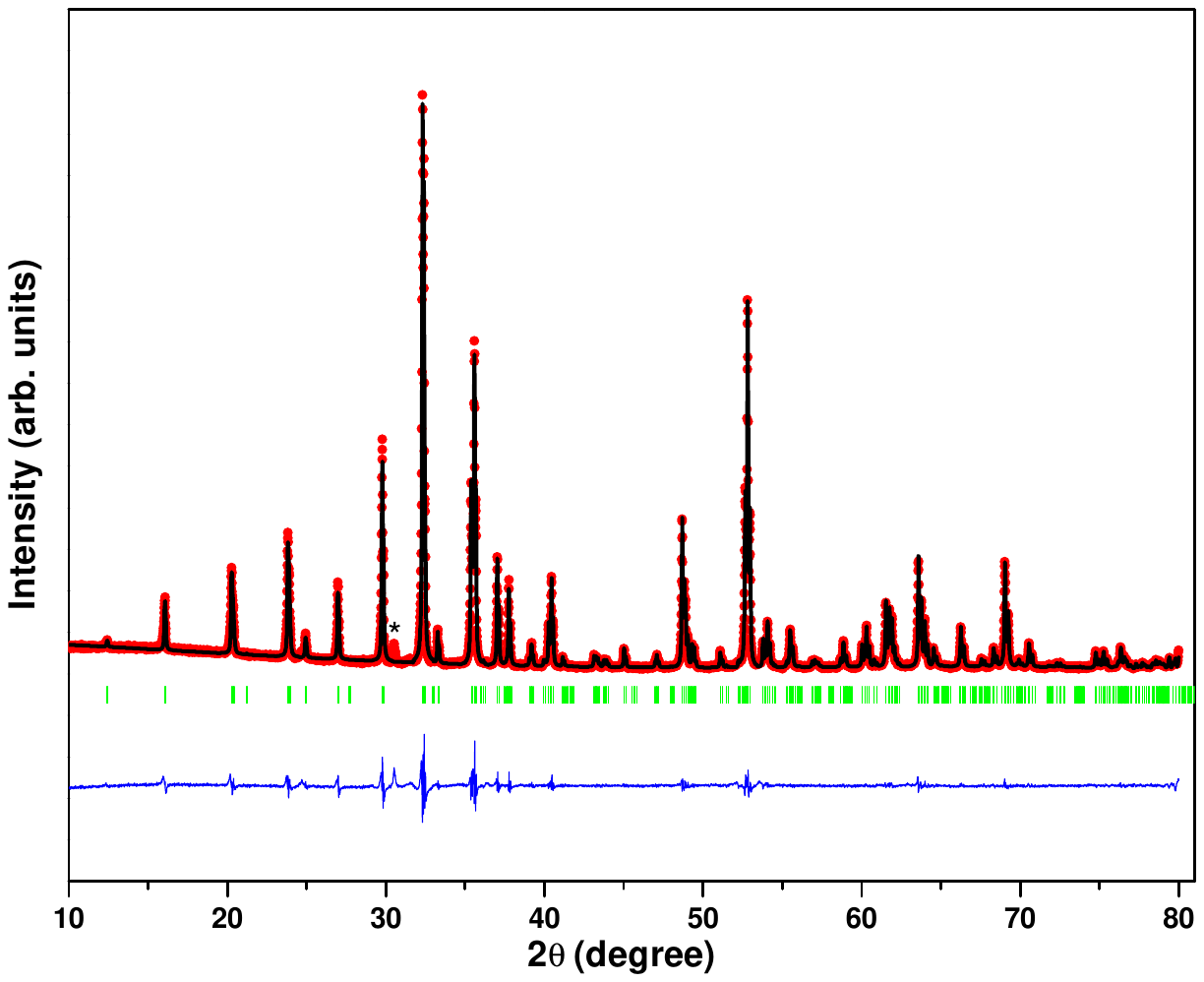}
\caption{\label{fig1} Rietveld refined powder x-ray diffraction patten of Ni$_4$Nb$_2$O$_9$ at ambient conditions recorded with Cu-K$_\alpha1,2$  radiation. A very weak peak at around 31$^{\circ}$ is marked by *. The solid red circles are the data points, the black line is the fit to the data, allowed peak positions are identified with green vertical ticks. The blue line highlights the difference between theoretical fit and experimental pattern.}
\end{figure}

\begin{figure}
\includegraphics[width=18cm]{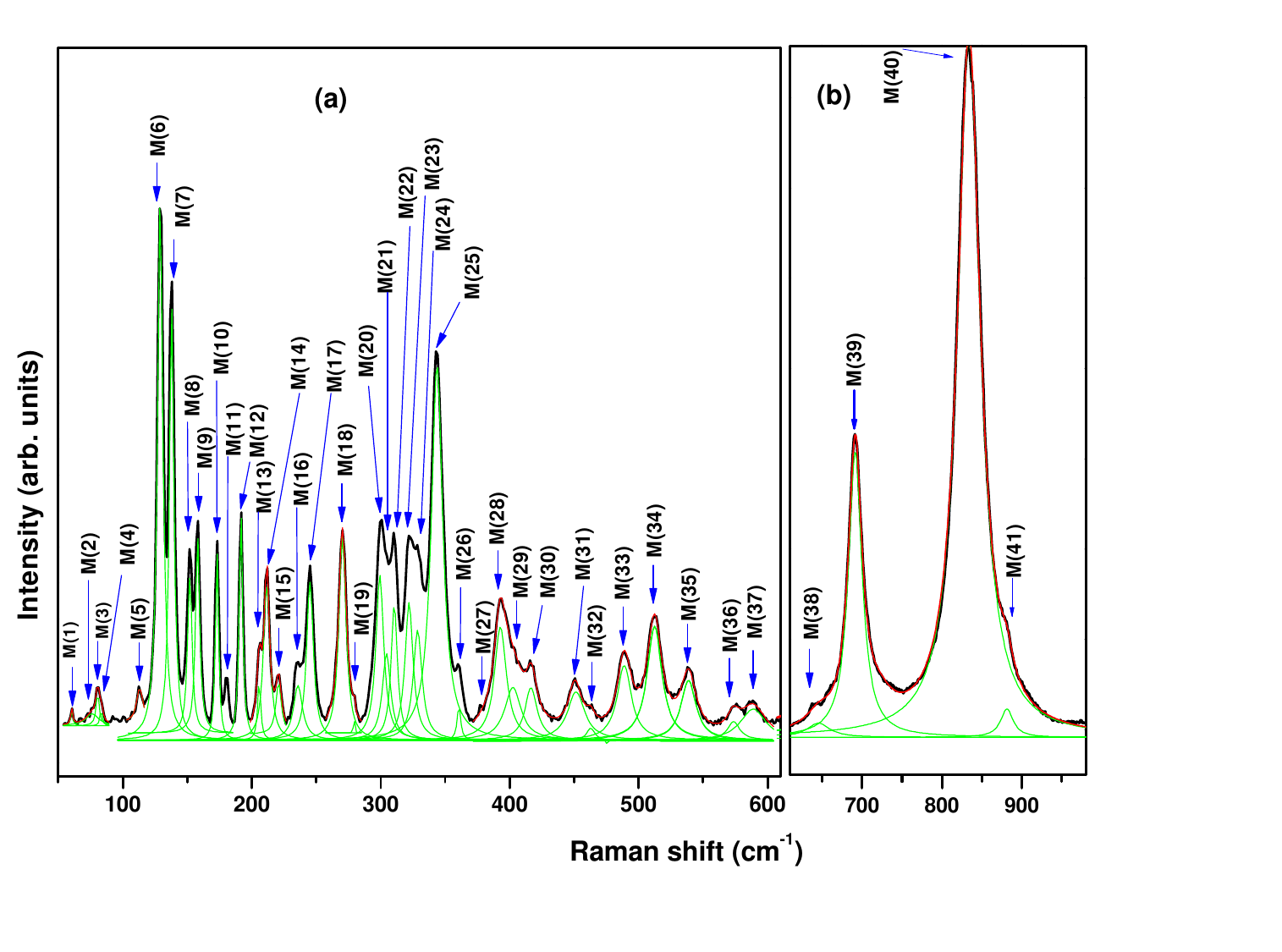}
\caption{\label{fig2} Ambient Raman spectrum of NNO along with Lorenzian profile fitting to the 41 Raman modes (a) 50-710 cm$^{-1}$; and (b) 610-980 cm$^{-1}$. The blue arrows identify the mode frequencies and labeled by M(1) to M(41).}
\end{figure}  

\newpage
\begin{figure}
\includegraphics[width=16cm]{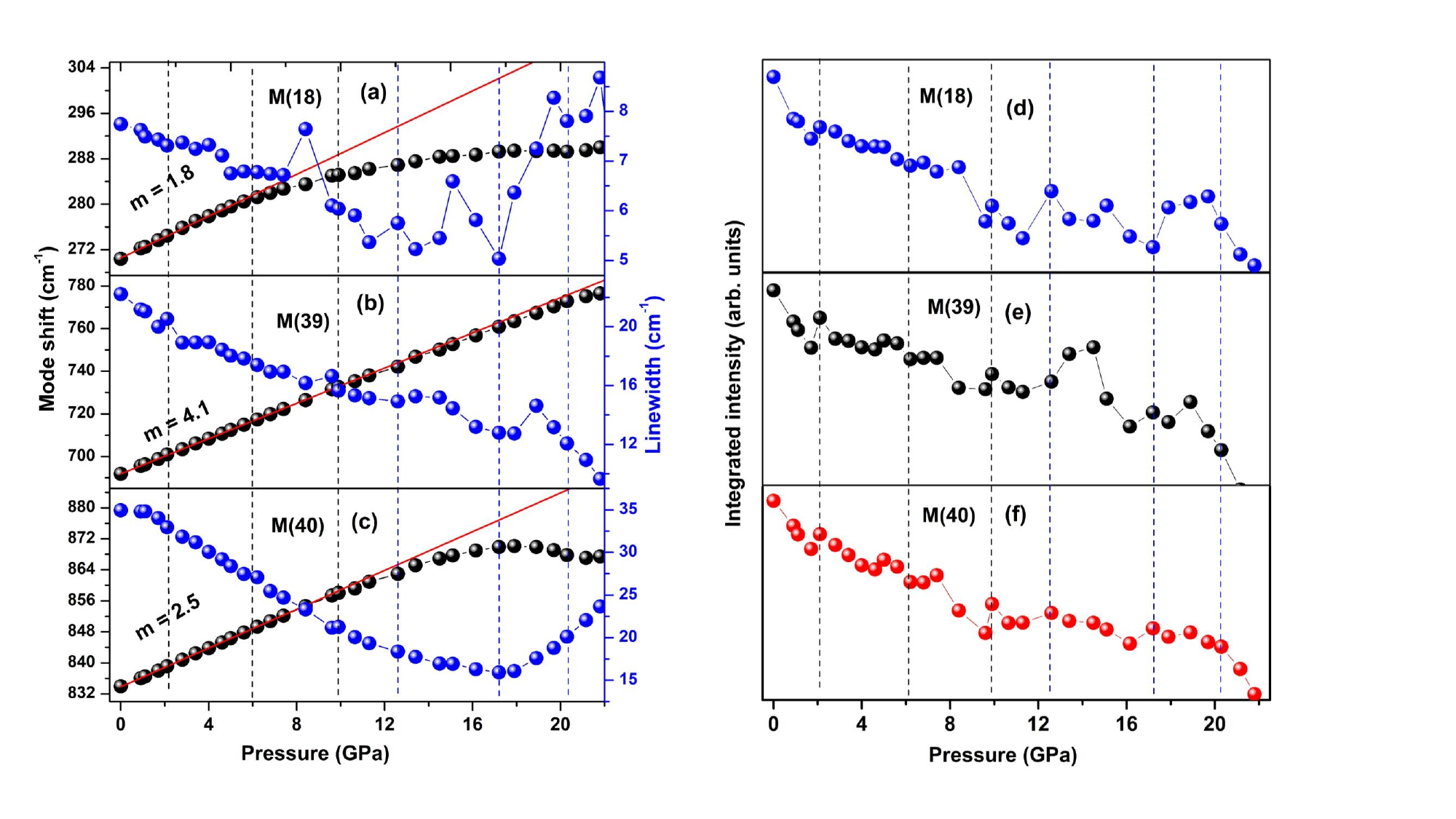}
\caption{\label{fig3} Pressure evolution of the frequency and linewidth for modes (a) M(18), (b) M(39), (c) M(40); alongside (d)–(f) the pressure dependence of their respective integrated intensities. Red solid lines represent linear fits to the mode frequencies. Black dashed lines highlight three isostructural transitions, while blue dashed lines mark pressure values corresponding to a long-range structural transition or major structural modifications.}
\end{figure}

\newpage
\begin{figure}
\includegraphics[width=8cm]{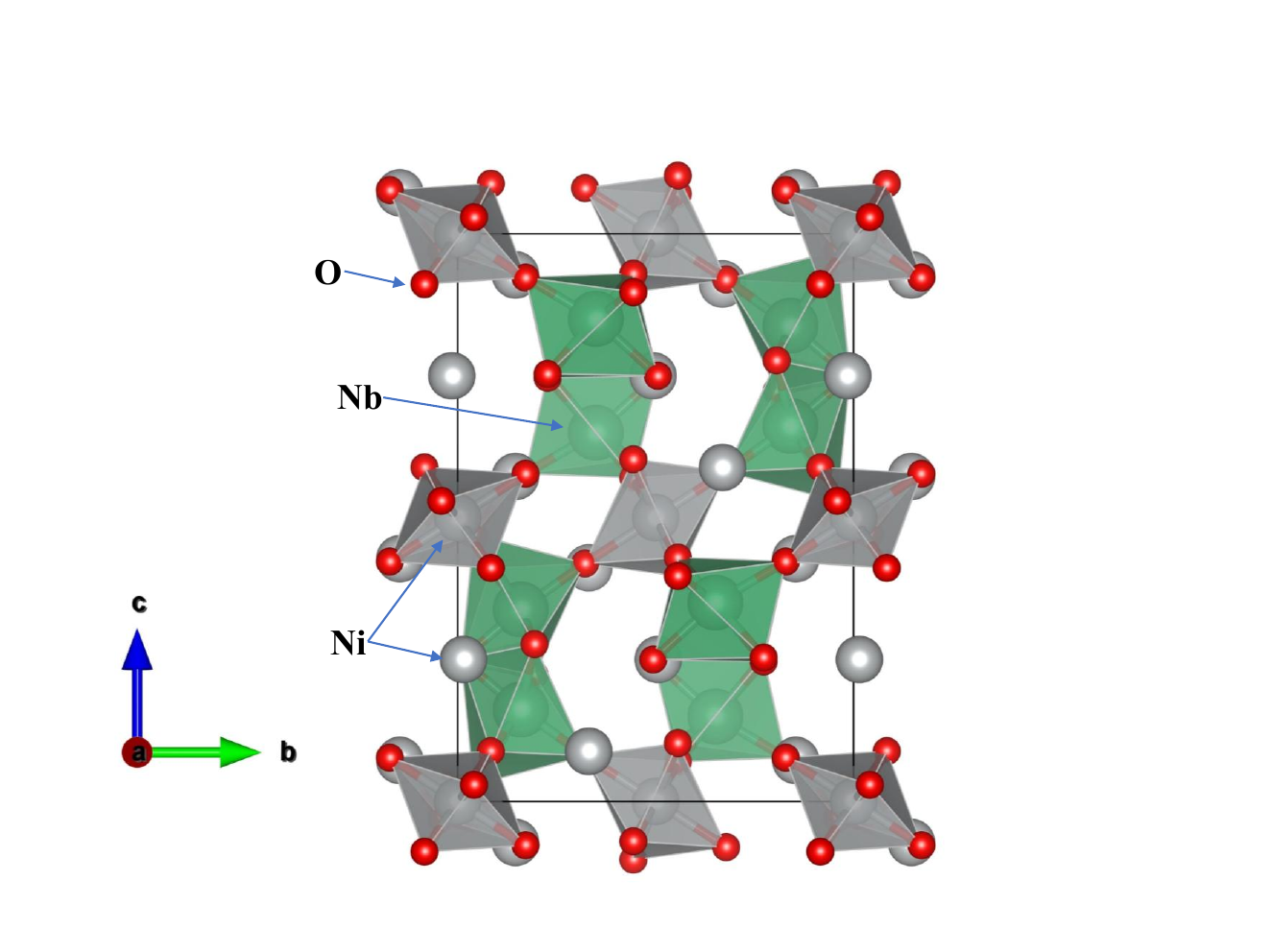}
\caption{\label{fig4} Structural illustration of the NNO unit cell at 34 GPa, crystallized in the monoclinic \textit{P2/c} symmetry.}
\end{figure} 

\newpage
\begin{table}[H]
\caption{\label{tab:tableS1
}
Rietveld refined  structural parameters of Ni$_4$Nb$_2$O$_9$ in orthorhombic \textit{Pbcn} (No. 60, Z = 4) symmetry with lattice parameters \textit{a} = 8.7188(4) \AA, \textit{b} = 5.0725(6) \AA, \textit{c} = 14.2921(5) \AA, at ambient pressure.}
\begin{ruledtabular}
\begin{tabular}{ccccccc}
 atom & site& x & y & z & U$_{iso}$ \\
\hline
 Ni1 & 8\textit{d} & 0.3301(5)  & 0.9921(2)  & 0.4993(3) & 0.0046(11)\\
 Ni2 & 8\textit{d} & 0.3399(4)  & -0.0138(6)  & 0.6903(2) & 0.0058(12)\\ 
 Nb &  8\textit{d} & 0.0287(5)  & -0.0082(6)  & 0.8553(4) & 0.0023(15)\\
 O1 & 4\textit{c} & 0  & 0.2827(7)  & 0.75 & 0.0014(13)\\
 O2 & 8d & 0.1674(9)  & 0.1723(11) & 0.9251(8) & 0.0015(16)\\
 O3 & 8\textit{d} & 0.1604(13)  & 0.1691(14) & 0.5897(15) & 0.0018(18)\\
 O4 & 8\textit{d} & 0.1461(16)  & 0.1425(15) & 0.2513(17) & 0.0013(17)\\
 O5 & 8\textit{d} & 0.4996(11)  & 0.1574(14) & 0.9094(12) & 0.0021(19)\\
\end{tabular}
\end{ruledtabular}
\end{table}

\newpage  
\begin{table*}[b]
\caption{\label{tab:tableS2
}
Rietveld refined crystal structural parameters of Ni$_4$Nb$_2$O$_9$ at 34 GPa in space group \textit{P2/c} (No. 13, Z = 4) with lattice parameters a = 5.7624(5), b = 8.8689(7), c 12.1898(3) \AA, $\beta$ = 91.65(5)$^\circ$.}
\begin{ruledtabular}
\begin{tabular}{cccccccccc}
  atom & site & x & y & z & U$_{iso}$ \\
\hline
 Ni1 &  4$g$ & 0.5046(5) & 0.1461(7) & 0.9277(4) & 0.0185(13) \\ 
 Ni2 & 2$e$ & 0 & 0.0147(7) & 1/4 & 0.0173(12) \\
 Ni3 & 2$f$ & 1/2 & 0.5065(5) & 1/4 & 0.0183(13) \\
 Ni4 & 4$g$ & 0.0098(8) & 0.6689(7) & 0.9117(5) & 0.0179(11) \\
 Ni5 & 2$a$ & 0 & 0 & 0 & 0.0214(16) \\ 
 Ni6 & 2$b$ & 1/2 & 1/2 & 0 & 0.022 (14) \\
 Nb1 & 4$g$ & 0.5004(4) & 0.1594(5) & 0.1616(7) & 0.0213(13) \\ 
 Nb2 & 4$g$ & 0.0051(8) & 0.6511(5) & 0.1496(6) & 0.0224(12) \\  
 O1 & 4$g$ & 0.7399(11) & 0.1942(12) & 0.2771(14) & 0.0218(13) \\  
 O2 & 4$g$ & 0.1521(13) & 0.7728(12) & 0.2535(15) & 0.0279(16) \\  
 O3 & 2$e$ & 0 & 0.5058(14) & 3/4 & 0.0226(11) \\
 O4 & 2$f$ & 1/2 & -0.0094(16) & 3/4 & 0.0231(15) \\
 O5 & 4$g$ & 0.2886(14) & 0.0839(14) & 0.4115(16) & 0.02251(14) \\
 O6 & 4$g$ & 0.8277(12) & 0.5562(15) & 0.4298(14) & 0.0252(13) \\
 O7 & 4$g$ & 0.0444(17) & 0.1714(12) & 0.92280(12) & 0.0243(14) \\
 O8 & 4$g$ & 0.4587(15) & 0.6782(16) &  0.9194(13) & 0.0236(15) \\
 O9 & 4$g$ & 0.6866(14) &  0.0412(15) & 0.4706(16) & 0.0224(14) \\ 
 O10 & 4$g$ & 0.2522(16) & 0.5562(14) & 0.3974(15) & 0.0212(16) \\
 
\end{tabular}
\end{ruledtabular}
\end{table*}
\end{document}